\documentclass[final,5p,times,twocolumn]{elsarticle}

\usepackage{enumitem}
\usepackage{url}
\usepackage{booktabs}
\usepackage{xspace}
\usepackage{setspace}
\usepackage[toc,title,page]{appendix}

\newcommand{\smat}[1]{\ensuremath{\left[\begin{smallmatrix}#1\end{smallmatrix}\right]}}

\newtheorem{definition}{Definition}

\newtheorem{corollary}{Corollary}
\newtheorem{theorem}{Theorem}
\newtheorem{proposition}{Proposition}

\newtheorem{assumption}{Assumption}
\newtheorem{example}{Example}

\graphicspath{{./figures/}}

\usepackage{graphicx,psfrag}
\usepackage{amsmath,amssymb,amsfonts,mathrsfs,mathtools}
\usepackage{color,epstopdf}
\usepackage{algorithmic}
\usepackage{enumitem}
\usepackage{subfigure}
\usepackage{floatrow}
\usepackage{float}
\usepackage{stfloats}
\usepackage{flushend}
\usepackage{bbold}
\usepackage{dsfont}
\usepackage{mathrsfs}
\usepackage{relsize}
\usepackage{hyperref}

\newtheorem{lem}{Lemma}

\newcommand{\R}{\mathbb{R}}

\newcommand{\1}{\mathbf{1}} 

\newcommand{\bbm}{\begin{bmatrix}}
\newcommand{\ebm}{\end{bmatrix}}

\def\qedp{\hspace*{\fill}~{\tiny $\blacksquare$}}

\def\be{\begin{equation}}
\def\ee{\end{equation}}
\def\ba{\begin{array}}
\def\ea{\end{array}}
\def\eqa{\begin{eqnarray}}
\def\eqe{\end{eqnarray}}

\definecolor{darkgreen}{rgb}{0.0, 0.55, 0.0}
\definecolor{amaranth}{rgb}{0.9, 0.17, 0.31}
\usepackage[prependcaption,colorinlistoftodos]{todonotes}

\def\qedp{\hspace*{\fill}~{\tiny $\blacksquare$}}
\def\qedd{\hspace*{\fill}~{\tiny $\Box$}}

\begin{document}

\begin{frontmatter}

\title{Data-driven 
{\color{black}harmonic} output regulation of {\color{black}a class of} nonlinear systems\tnoteref{t1}}
\tnotetext[t1]{This publication is part
of the project Digital Twin with project number P18-03 of
the research programme TTW Perspective which is (partly)
financed by the Dutch Research Council (NWO). The first author is supported by the China Scholarship Council.}
\tnotetext[]{Corresponding author: Zhongjie Hu.
\author{Zhongjie Hu, Claudio De Persis}
\ead{\{zhongjie.hu,c.de.persis\}@rug.nl}
\author{{\color{black}John W.~Simpson-Porco}}
\ead{jwsimpson@ece.utoronto.ca}
\author{Pietro Tesi}
\ead{pietro.tesi@unifi.it}
}

\author[1]{Zhongjie Hu, Claudio De Persis} 
\affiliation[1]{organization={Engineering and Technology Institute Groningen (ENTEG)},
            addressline={University of Groningen}, 
            postcode={9747 AG Groningen}, 
            country={The Netherlands}}

\author[2]{{\color{black}John W.~Simpson-Porco}} 
\affiliation[2]{organization={Department of Electrical and Computer Engineering},
            addressline={University of Toronto}, 
            city={Toronto},
            state={ON},
            country={Canada}}

\author[3]{Pietro Tesi} 
\affiliation[3]{organization={DINFO},
            addressline={University of Florence}, 
            postcode={50139 Florence}, 
            country={Italy}}

\begin{abstract}
The paper deals with the data-based design of {\color{black} state-feedback} controllers that solve the output regulation problem for {\color{black}a class of} nonlinear systems. Inspired by recent developments in  model-based output regulation design techniques and in data-driven control design for nonlinear systems, we derive a data-dependent semidefinite program that, when solved, directly returns a controller that 
{\color{black}steers the regulation error to a periodic signal 
whose Fourier series has identically zero coefficients up to a certain order set by the controller.} When specialized to the case of linear systems, the result appears to improve upon existing work. Numerical results illustrate the findings. 
\end{abstract}

\begin{keyword}
Data-driven control \sep nonlinear systems
\end{keyword}

\end{frontmatter}


\section{Introduction}
\label{sec:intro}
The last few years  have witnessed a concerted effort to use data collected in offline experiments for {\em directly} designing control laws for uncertain dynamical systems. The qualifier ``direct" refers to the fact that such an approach has focused on deriving data-dependent convex programs that, when solved, immediately return a controller with certified guarantees of correctness. 
{\color{black} After  numerous works  pioneered the direct data-driven design for linear control systems}, more recent efforts have concentrated on nonlinear control systems -- see \cite{dpt2023arc,martin2023guarantees}. 

In one of these recent results for nonlinear control systems \cite{hu2024enforcing}, we have explored the role of contractivity (\cite{lohmiller1998contraction,pavlov2004convergent,sontag2010contractive,FB-CTDS}) for designing controllers from data for nonlinear systems. A benefit of using contractivity is that it allows the designer to 
address significant control problems beyond stabilization, one of these being the output regulation problem  
\cite{pavlov2006uniform,giaccagli2022sufficient,giaccagli2023lmi}. This feature was exploited in \cite{hu2024enforcing} to derive  a data-dependent semidefinite program (SDP) for tuning nonlinear integral controllers that guarantee  tracking of constant reference signals and rejection of constant disturbances in nonlinear systems {\color{black} of the same class as the one considered here}. 

In this note, we extend the result of \cite{hu2024enforcing} to the case of periodic exogenous signals. Since we are  interested in designing output regulators for unknown nonlinear systems, we refrain from assuming the existence of a normal form of the system, which is not available to the designer. {\color{black}We focus on a class of nonlinear systems whose drift vector field can be expressed as a linear combination of elements in a library of functions and whose input vector field is constant. We}  
approach the problem by 
{\color{black} augmenting the plant with a post-processing internal model (adopting the terminology in \cite{astolfi2013nonlinear}),}
and then designing a ``stabilizer" {\color{black} from data} that {\color{black} renders the resulting closed-loop system} contractive. Contractivity of such {\color{black}a} closed-loop system guarantees that the steady-state response of the system is also periodic, and this property is used to conclude that 
{\color{black}
certain harmonics of the steady state regulation error are zero.
}

In a model-based context, the approach outlined above has been thoroughly investigated. 
Adding the internal model first and then the stabilizer, and using Fourier series properties to infer 
{\color{black} that some of the  coefficients of the  Fourier series of the error vanish}
has been inspired by previous works (\cite{astolfi2015approximate,astolfi2022harmonic,giaccagli2024incremental}).  
{\color{black} 
The results in \cite[Theorem 1]{astolfi2015approximate} also show how to make the regulation error arbitrarily small in the $\mathcal{L}_2$-sense for disturbances of sufficiently small magnitude. {\color{black} See  \cite[Section 2.5]{byrnes1997output} for an early formulation of an approximate output regulation problem.}} 

The main novelty of this paper is to use the data-driven control design framework of \cite{de2019formulas,dprt2023cancellation,hu2024enforcing} to reduce the design of  output regulators for 
{\color{black} a class of} unknown nonlinear systems to the solution of a data-dependent semidefinite program. To achieve this goal, in addition to the tools in the above mentioned references, we establish a new technical result that filters out the impact of the exogenous signals on the data-dependent representation used for the analysis. When specialized to the case of linear systems, the result {\color{black} provides a methodological improvement over existing work, in the sense that no measurement of the disturbance is needed to design the controller.}
{\color{black} On the other hand, we do not provide any results on the design of controllers that make the $\mathcal{L}_2$-norm of the steady state regulation error arbitrarily small.}

{\it Outline.} We introduce the {\color{black} class of} uncertain nonlinear systems of interest and recall about the output regulation problem in Section \ref{sec:sys-interest}. The dataset used for the design and the important technical lemma (Lemma \ref{lem:forma-nuova-risposta-esosistema}) that leads to a perturbation-free data-dependent representation are discussed in Section \ref{sec:data-depend-repres}. The correctness of the regulator constructed from data is proven in Section \ref{sec:db-design-regulator}. In Section \ref{sec:linear-systems} we discuss the special case of linear systems. An appendix contains the proof of Lemma \ref{lem:forma-nuova-risposta-esosistema}, a few facts about the Fourier series and an interesting technical result from \cite{astolfi2015approximate}.

\section{Systems of interest and the output regulation problem
} \label{sec:sys-interest}
Consider the controlled plant
\be\label{controlled.plant.data}
\ba{rl}
\dot w=& Sw\\
\dot x =& AZ(x) +Bu +Pw\\
y =& 
CZ(x)+Qw
\ea
\ee
where $w\in \R^{n_w}$, 
$x\in \R^n$, $u\in \R^m$. We assume that the matrix $S\in \mathbb{R}^{n_w\times n_w}$ is a known matrix, while the signal $w$ is not measured.  As in \cite{dprt2023cancellation}, \cite{hu2024enforcing},  we express the drift vector field as a linear combination of known functions in $Z\colon \R^n \to \R^{q}$, which includes linear and nonlinear functions,  namely, $Z(x)=\begin{bmatrix}
x\\ Q(x)
\end{bmatrix}$, where  $Q: \mathbb R^n \rightarrow \mathbb R^{q-n}$ contains only nonlinear functions. All the functions of $Z(x)$ are continuously differentiable.  
{\color{black} The functions in $Z(x)$ constitute the library of functions through which the drift vector field can be expressed. 
Expressing the drift vector field $f(x)$ as $AZ(x)$ is customary in sparse identification \cite{trajectory-matrix-application} and results in no loss of generality if each component of $f(x)$ lies in the span of the elements of the library $Z(x)$. Otherwise, neglected nonlinearities  or remainders should be taken into account as it is done in \cite[Section VI.B]{dprt2023cancellation} and \cite{hu2023learning}. Similarly, considering constant input vector fields could be relaxed following \cite[Section VI.B]{dprt2023cancellation}. However, these relaxations would complicate considerably the analysis and are not further  pursued here.  
}

{\color{black}In \eqref{controlled.plant.data},} the matrices $A,B,P$ are unknown. The matrices $C$ and $Q$ are partially unknown, as we clarify 
{\color{black} next. First,} the vector $y$ is the vector of measured outputs and is split as 
\be\label{y.data}
y =\begin{bmatrix}
e\\
y_r
\end{bmatrix}
=
\begin{bmatrix}
C_e Z(x)+Q_e w\\
C_r Z(x)+Q_r w
\end{bmatrix}
\ee
where $e\in \R^p$ is the vector of error variables and $y_r\in \R^{p_r}$ contains the other measured variables in addition to $e$. Here, we consider the case in which the state $x$ is fully measured, then $y_r=x$, i.e.,
\[
C_r = \begin{bmatrix} I_n & 0_{n\times (q-n)}\end{bmatrix}, \quad Q_r=0.
\]
As a consequence, $x,e$ will be both available to the controller during the execution of the control task.  On the other hand, $C_e$ and $Q_e$ are unknown matrices.

\smallskip

As is customary in {\color{black} post-processing output regulation design}, an internal model of the exosystem of the form
\be \label{lin-outp-regulator-nl}
\dot \eta = \Phi \eta + Ge
\ee
is added, where the pair of matrices $\Phi, G$ will be specified in \eqref{d-astolfi-im}-\eqref{d-astolfi-im2} below. 
{\color{black}Following the terminology in \cite{giaccagli2024incremental},} we will solve 
{\color{black}a harmonic} 
regulation problem 
using data to offset the uncertainty about the model \eqref{controlled.plant.data}. 
By solving 
{\color{black}a harmonic 
regulation problem using data}, it is meant designing a controller {\color{black} from data} which guarantees the regulation error of the closed-loop system converges to a periodic signal {\color{black}whose Fourier series has identically zero coefficients up to a certain order}
(see Theorem \ref{cor:integr-control} below for a formal statement). 
{\color{black} Before proceeding}, it is useful to recall the {\color{black} (global)} output regulation problem. This is  the problem of designing a controller $u =  k(x,\eta)$ such that, {\color{black} for any initial condition, 
the solution} 
of the closed-loop system   
\begin{subequations}
\label{plant-nl}
\begin{alignat}{3}
\dot w=&Sw \label{plant1-nl}
\\
\dot x =& AZ(x) +Bk(x,\eta) +Pw\label{plant2-nl}
\\
\dot \eta = & \Phi\eta+Ge
\\
\begin{bmatrix}  
e\\
x\\
\eta
\end{bmatrix} 
=
& 
\begin{bmatrix}  
\begin{bmatrix}
\multicolumn{2}{c}{C_e} \\ 
I_n & 0_{n\times (q-n)}
\end{bmatrix} 
& 0\\
0 & I
\end{bmatrix}
\begin{bmatrix}  
Z(x) \\
\eta
\end{bmatrix}
+
\begin{bmatrix}  
\begin{bmatrix}
Q_e \\ 0
\end{bmatrix} \\
0
\end{bmatrix}w\label{plant3-nl}
\end{alignat}
\end{subequations}
satisfies (i)  
$x(t), \eta(t)$ are bounded for all $t\ge 0$ and (ii) $e(t)\to 0$ as $t\to +\infty$.

\smallskip

To tackle the 
{\color{black} harmonic}
output regulation problem, we first restrict the exosystem.
\begin{assumption}\label{assum:w}
For any initial condition $w(0)$, the solution $w(t)$  of the exosystem $\dot w = Sw$ passing through $w(0)$ exists for all 
{\color{black} $t\in \R$}
and is bounded. Moreover, it is a $D$-periodic function, for some period $D$.     
\end{assumption}

{\color{black}Assumption \ref{assum:w} is equivalent to assuming that the matrix $S$ has semisimple eigenvalues, all lying on the imaginary axis, and whose imaginary parts are rationally related\footnote{The positive real numbers $\lambda_1, \ldots, \lambda_\rho$, for some integer $\rho\ge 2$, are said to be rationally related if there exists a positive real number $\Lambda$ such that $\lambda_i = \ell_i \Lambda$ for positive integers $\ell_i$, $i=1,2,\ldots, \rho$.}. Thus, it is slightly more restrictive than assuming that the exosystem is a neutrally stable linear system. In some applications, the signal $w$ may model both an \emph{unknown} disturbance signal $d$ and a \emph{desired} reference signal $r$, i.e., $w=\left[d^\top \; r^\top \right]^\top$. In this case, one may further relax Assumption \ref{assum:w} above by assuming that $d$ is generated by a linear exosystem,  say $\dot d=S_d d$, where $S_d$ is such that the solution $d(t)$ satisfies Assumption \ref{assum:w}, while $r(t)$ exists for all 
$t\in \R$,
is bounded, $D$-periodic and can be measured, but is not necessarily generated by a finite-dimensional linear exosystem. This relaxation will be discussed in Section \ref{subsec:rel-Asspt1}.} 

As for the internal model controller  \eqref{lin-outp-regulator-nl}, we follow \cite{astolfi2022harmonic} and consider the matrices $\Phi, G$ defined as 
\be\label{d-astolfi-im}
\Phi = {\rm block.diag}(\underbrace{\phi, \ldots, \phi}_{p \text{ times}}) , \quad G={\rm block.diag}(\underbrace{\Gamma, \ldots, \Gamma}_{p \text{ times}})
\ee
where 
\be\label{d-astolfi-im2}
\ba{l}
\phi = {\rm block.diag}(0, \phi_1,  \ldots, \phi_\ell), 
 \phi_k = \begin{bmatrix}
0 & \omega_k\\
-\omega_k & 0 
\end{bmatrix}\!, \omega_k=k \frac{2\pi}{D}\\
\Gamma = {\rm col} (\gamma, N, \ldots, N), \quad \gamma \in \R\setminus  \{0\},\quad  N\in \mathbb{R}^2\setminus \{0\}. 
\ea\ee
The internal model incorporates an integrator and $\ell$ harmonic oscillators that generate the first $\ell$ harmonics whose frequencies are multiple of the fundamental frequency $\frac{2\pi}{D}$. {\color{black} These harmonics are included because the presence of nonlinearities in the plant may generate such harmonics in the error signal, which must in turn be compensated for by the internal model.} The number of harmonics $\ell$ to be embedded in the internal model is a design parameter that will be discussed later. 
For $k=1,2, \ldots, p$, the $k$-th subsystem of the internal model is of the form
\be\label{im-k}
\dot \eta_k = \begin{bmatrix}
0 & 0  & \ldots & 0\\
0  & \phi_1 & \ldots  & 0\\  
\vdots  & \vdots & \ddots & \vdots \\
 0 & 0 & \ldots &  \phi_\ell \end{bmatrix}
 \eta_k +
 \begin{bmatrix} \gamma \\ N\\ \vdots\\ N\end{bmatrix} e_k 
\ee
where $\eta_k \in \R^{2\ell+1}$, $e_k\in \R$ is the $k$-th component of $e\in \R^p$. Hence, $\eta\in \R^{n_\eta}$, where  $n_\eta:= p(2\ell+1)$.

\section{The dataset and a data-dependent representation}\label{sec:data-depend-repres}
To ease the presentation,  we rearrange the vectors $x$, $Q(x)$, $\eta$ {\color{black}into} a new vector {\color{black} -- see Eq.\eqref{calZ} below}. First, we partition the matrices $A,C_e$ in 
\eqref{plant-nl} according to the structure of $Z(x)$ as
\[
\begin{bmatrix}
A \\
C
\end{bmatrix}
=
\begin{bmatrix}
\,\overline A &  \widehat A\\
\,\overline C_e & \widehat C_e
\end{bmatrix}
\]
and stack  $Z(x)$ and $\eta$, the state of the internal model \eqref{lin-outp-regulator-nl}, in the vector 
\be\label{calZ}
\mathcal{Z}(x, \eta)=
\begin{bmatrix}
x\\
\eta\\
Q(x)
\end{bmatrix}.
\ee
Having introduced the internal model \eqref{lin-outp-regulator-nl}, \eqref{d-astolfi-im}, \eqref{d-astolfi-im2}, we write the resulting augmented system as 
 \begin{subequations}
\label{augm-sys}
\begin{alignat}{2}
\begin{bmatrix}  
\dot x \\
\dot \eta
\end{bmatrix}
=& 
\begin{bmatrix}
\,\overline A & 0_{n\times n_\eta} & \widehat A\\
\,G\overline C_e & \Phi & G \widehat C_e
\end{bmatrix}
\mathcal{Z}(x, \eta)
+
\begin{bmatrix}  
B \\
0
\end{bmatrix}
u
+
\begin{bmatrix}  
P \\
GQ_e
\end{bmatrix}
w\label{augm-sys1}
\\
:= &\mathcal{A} 
\mathcal{Z}(x, \eta)
+
 \mathcal{B} u
 +
\mathcal{P} w, \label{augm-sys2}\\
y_a := & 
\begin{bmatrix}  
e\\
x\\
\eta
\end{bmatrix} 
\!=\! 
\begingroup 
\setlength\arraycolsep{1.8pt}
\begin{bmatrix}  
\; \overline{C}_e & 0_{p\times n_\eta} & \widehat C_e\\ 
I_n & 0_{n\times n_\eta} & 0_{n\times (q-n)}\\
0_{n_\eta\times n} & I_{n_\eta} & 0_{n_\eta\times (q-n)}
\end{bmatrix}
\endgroup
\mathcal{Z}(x,\eta)
+ 
\begin{bmatrix}  
Q_e \\ 0
\\
0
\end{bmatrix}w.
\label{augm-sys3}
\end{alignat}
\end{subequations}
Note that the vector  $y_a$ of  measured outputs of the augmented system includes both $y$ defined in \eqref{y.data} and $\eta$. 

If we run an experiment on the augmented system  \eqref{augm-sys} and consider $T$ samples of 
the output $y_a$, input $u$ and exosystem state $w$, we obtain the relation
\be\label{data-matrices-identity}
\mathcal{Z}_{1} = \mathcal{A}  \mathcal{Z}_{0} +\mathcal{B} U_0 +\mathcal{P} W_0
\ee
where
\begin{subequations}
\label{eq:data}
\begin{align}
U_0 &:=\begin{bmatrix} u (t_0) &\ldots & u(t_{T-1})\end{bmatrix} 
 \\
 \mathcal{Z}_0 &:= \begin{bmatrix}
x(t_0) &  x(t_1) & \ldots &  x(t_{T-1})\\ 
\eta(t_0)&  \eta(t_1) & \ldots & \eta(t_{T-1})\\ 
Q(x(t_0)) &  Q(x(t_1)) & \ldots &  Q(x(t_{T-1}))\\ 
\end{bmatrix}  \\
 \mathcal{Z}_1 &:= \begin{bmatrix} \dot x(t_0) &\ldots & \dot x(t_{T-1}) \\ \dot \eta(t_0) &\ldots & \dot \eta(t_{T-1})
\end{bmatrix} 
 \\ 
W_0 &:= \begin{bmatrix} w (t_0) &\ldots & w(t_{T-1}) \label{eq:data4}
\end{bmatrix}   
\end{align}
\end{subequations}
and  $0<t_0<\ldots<t_{T-1}$. 
The samples of $\dot \eta$ are computed via the samples of $\eta$ and $e$ based on \eqref{lin-outp-regulator-nl}. {\color{black} In fact, as the internal model \eqref{lin-outp-regulator-nl} is a part of the control design, one may use the dataset generated by the experiment on the physical plant to collect the samples of $\eta$ and $\dot \eta$ in a distinct phase. }The vector $w$ may model unmeasured disturbances, and hence the effect cannot be excluded from the plant's dynamics during the data acquisition phase. Hence, the dataset is perturbed by the unmeasured samples of $w$ collected in $W_0$. For this reason, while the matrices of data $\mathcal{Z}_{0},\mathcal{Z}_{1},  U_0$ are available to the designer,  the matrix $W_0$ is not. Nonetheless, the following {\color{black} new technical} result, which extends the arguments in \cite[Section VI]{hu2024enforcing},  underscores that the matrix  $W_0$ can be factored into a product of two matrices, one of which is known, 
allowing us -- as we will see -- to design the control law without {\color{black} requiring the samples $W_0$ of $w$}. {\color{black}We remark that, for the sake of generality, we state the result in a slightly more comprehensive form than needed for our purposes in this article and without requiring $S$ to satisfy Assumption   \ref{assum:w}.}

\begin{lem}\label{lem:forma-nuova-risposta-esosistema}
Consider the system 
\[
\dot w= Sw, 
\]
with initial condition $w(0)$ and 
$w\in \R^{n_w}$. Let 
$\mathcal{T}\in \R^{n_w\times n_w}$ 
 be the similarity matrix that transforms $S$ into its real Jordan form $J$, namely $J= \mathcal{T}^{-1}S\mathcal{T}$. Then there exist positive
 integers $k_1, \ldots, k_{r+s}$ satisfying $k_1+\ldots+k_r+2k_{r+1}+\ldots +2k_{r+s}=n_w,$  matrices  $L_{k_i}(w(0))\in \R^{k_i\times k_i}$, $M_{k_i}(\lambda_i, t)\colon \mathbb{R}_{\ge 0}$ $\to  \mathbb{R}^{k_i}$ associated with the real eigenvalues $\lambda_i$ and matrices $L_{k_{r+i}}(w(0))\in \R^{2 k_{r+i}\times 2 k_{r+i}}$, $M_{k_{r+i}}(\lambda_{r+i},  \lambda_{r+i}^\star,t)\colon \mathbb{R}_{\ge 0}\to \R^{2 k_{r+i}}$ associated with  complex conjugate eigenvalues $\lambda_{r+i}=\mu_i+\sqrt{-1}\psi_i,  \lambda_{r+i}^\star=\mu_i-\sqrt{-1}\psi_i$ 
such that 
\[
w(t) = \mathcal{T} \begin{bmatrix}
L_{k_1}(w(0)) M_{k_1}(\lambda_1, t)\\
\vdots\\
L_{k_r}(w(0)) M_{k_r}(\lambda_r, t)\\
L_{k_{r+1}}(w(0)) M_{k_{r+1}}(\lambda_{r+1},  \lambda_{r+1}^\star,t)\\
\vdots\\
L_{k_{r+s}}(w(0)) M_{k_{r+s}}(\lambda_{r+s},  \lambda_{r+s}^\star,t)\\
\end{bmatrix},\quad \forall t\in \R_{\ge 0}.
\]
Furthermore, the matrices $M_{k_i}(\lambda_i, t)$, $M_{k_{r+i}}(\lambda_{r+i},  \lambda_{r+i}^\star,t)$ have the form 
\be\label{columns-M}
\ba{l}
\hspace{1.5cm}M_{k_i}(\lambda_i, t)
=
\begin{bmatrix}
\frac{t^{k_i-1}}{(k_i-1)!}{\rm e}^{\lambda_i t}\\[1mm]
\frac{t^{k_i-2}}{(k_i-2)!}{\rm e}^{\lambda_i t}\\[1mm]
\vdots\\
t {\rm e}^{\lambda_i t}\\[1mm]
{\rm e}^{\lambda_i t}
\end{bmatrix},
\\[16mm]
M_{k_{r+i}}(\lambda_{r+i},  \lambda_{r+i}^\star,t)
=
\begin{bmatrix}
\frac{t^{k_{r+i}-1}}{(k_{r+i}-1)!}  {\rm e}^{\mu_i t}\cos(\psi_i t)\\[1mm]
\frac{t^{k_{r+i}-1}}{(k_{r+i}-1)!}  {\rm e}^{\mu_i t}\sin(\psi_i t)\\[1mm]
\hline
\frac{t^{k_{r+i}-2}}{(k_{r+i}-2)!} {\rm e}^{\mu_i t}\cos(\psi_i t)\\[1mm]
\frac{t^{k_{r+i}-2}}{(k_{r+i}-2)!} {\rm e}^{\mu_i t }\sin(\psi_i t)\\[1mm]
\hline
\vdots\\
\hline
{\rm e}^{\mu_i t}\cos(\psi_i t)\\[1mm]
{\rm e}^{\mu_i  t}\sin(\psi_i t)
\end{bmatrix}.
\hspace*{\fill}~{\tiny \text{$\square$}}
\ea
\ee 
  
\end{lem} 
 
{\em Proof.} See \ref{app:proof}. \qedp

\begin{figure*}
\begin{subequations}
\label{LM}
\begin{align}
L&:= {\rm block.diag} (L_{k_1}(w(0)), \ldots, L_{k_r}(w(0)), L_{k_{r+1}}(w(0)), \ldots,  L_{k_{r+s}}(w(0))) \label{matrix-L} 
\\
\label{matrix-M} 
M&:=
\begin{bmatrix} 
M_{k_1}(\lambda_1, t_0) & M_{k_1}(\lambda_1, t_1) & \ldots & M_{k_1}(\lambda_1, t_{T-1})\\
\vdots\\
M_{k_r}(\lambda_r, t_0) & M_{k_r}(\lambda_r, t_1) & \ldots & M_{k_r}(\lambda_r, t_{T-1}) \\
M_{k_{r+1}}(\lambda_{r+1},  \lambda_{r+1}^\star,t_0) & M_{k_{r+1}}(\lambda_{r+1},  \lambda_{r+1}^\star,t_1) & \ldots & M_{k_{r+1}}(\lambda_{r+1},  \lambda_{r+1}^\star,t_{T-1})\\
\vdots &\vdots &\vdots &\vdots \\
M_{k_{r+s}}(\lambda_{r+s},  \lambda_{r+s}^\star,t_0) & M_{k_{r+s}}(\lambda_{r+s},  \lambda_{r+s}^\star,t_1) & \ldots & M_{k_{r+s}}(\lambda_{r+s},  \lambda_{r+s}^\star,t_{T-1})
\end{bmatrix}
\end{align}
\hrulefill
\end{subequations}
\end{figure*}

{\color{black}The result is based on expressing the solution of the exosystem via a {\color{black} real} Jordan form and then rearranging it to have the unknown terms depending on the initial conditions to appear on the left of the expression. This is the key to obtain a data-based representation of the closed-loop system that is independent of the unmeasured samples of $w$, as we illustrate next, and gives some distinctive advantages in the design of the controller.}  {\color{black} We also remark that if $S$ is restricted to satisfy Assumption \ref{assum:w} then the result holds with the positive integers $k_1, \ldots, k_{r+s}$ all equal to $1$. }

\medskip
Define the matrices $L$ and $M$ as in \eqref{LM}. We note that, by Lemma \ref{lem:forma-nuova-risposta-esosistema}, the factorization
\be\label{W0-identity}
W_0 = \mathcal{T} LM
\ee
of the matrix $W_0$ defined in \eqref{eq:data4} holds, where {\color{black} $L$ depends on the initial conditions of the exosystem and is therefore unknown, whereas $M$ depends on the Jordan decomposition of $S$ and as such} is {\em known}. In particular, since the matrix $W_0$ satisfies the identity \eqref{W0-identity},  we can rewrite the relation \eqref{data-matrices-identity} as 
\[
\mathcal{Z}_{1} = \mathcal{A}  \mathcal{Z}_{0} +\mathcal{B} U_0 +\mathcal{P} \mathcal{T}  LM.
\]
The following result, {\color{black}the importance of which we will comment on after Corollary \ref{cor:dd-representation} below,} is a straightforward extension of \cite[Lemma 3]{hu2024enforcing}, and hence its proof is omitted.
\begin{lem}\label{lem:dd-representation}
Consider any matrices $\mathcal{K} \in \mathbb R^{m \times (q+n_\eta)}$,
$\mathcal{G} \in \mathbb R^{T \times (q+n_\eta)}$ such that 
\begin{equation} \label{eq:GK-exosystem}
\begin{bmatrix} \mathcal{K} \\ I_{q+n_\eta} \\ 0_{n_w \times (q+n_\eta)}\end{bmatrix} = \begin{bmatrix} U_0 \\ \mathcal{Z}_0 \\ M\end{bmatrix} \mathcal{G}
\end{equation}
 holds. 
Then system \eqref{augm-sys2} under the control law 
\be\label{state-feedback}
u=\mathcal{K} {\color{black}\mathcal{Z}(x, \eta)}
\ee
results in
the closed-loop dynamics 
\begin{equation} \label{eq:GK_closed_noise_periodic}
\!\!\!\!\!\!
\ba{rl}
\begin{bmatrix}  
\dot x \\
\dot \eta
\end{bmatrix}
=& \mathcal{Z}_1 \mathcal{G} {\color{black}\mathcal{Z}(x, \eta)}
+
\mathcal{P}w.
\text{\hspace*{\fill}~{\tiny $\square$}}
\ea\end{equation}
\end{lem}

{\color{black} Before commenting on the main consequence of this result, we develop it a bit further.}
It was observed in \cite[Theorem 4]{hu2024enforcing} that there is redundancy in the condition \eqref{eq:GK-exosystem}, more specifically in the constraint $0=M\mathcal{G}$. In fact, consider the expression \eqref{matrix-M} of the matrix $M$ and those of its columns (cf.~\eqref{columns-M}). For any integer $0\le \tau \le T-1$, consider the expression of the vectors $M_{k_i}(\lambda_i, t_\tau)$ and $M_{k_j}(\lambda_j, t_\tau)$ associated with the same real eigenvalue, i.e.~$\lambda_i=\lambda_j$, but with different order of the corresponding Jordan blocks, i.e.~$k_i\ne k_j$. Let us assume without loss of generality that $k_i$ is the largest order associated with the eigenvalue  $\lambda_i=\lambda_j$. Then we see that the matrix made of the last $k_j$ rows of the submatrix 
\[
\begin{bmatrix} 
M_{k_i}(\lambda_i, t_0) & M_{k_i}(\lambda_i, t_1) & \ldots & M_{k_i}(\lambda_i, t_{T-1})
\end{bmatrix}
\]
of $M$ coincide with  
\[
\begin{bmatrix} 
M_{k_j}(\lambda_j, t_0) & M_{k_j}(\lambda_j, t_1) & \ldots & M_{k_j}(\lambda_j, t_{T-1})
\end{bmatrix}.
\]
Hence the constraint 
\[
{\scriptsize\begin{bmatrix} 
M_{k_j}(\lambda_j, t_0) & M_{k_j}(\lambda_j, t_1) & \ldots & M_{k_j}(\lambda_j, t_{T-1})
\end{bmatrix}}\mathcal{G}=0_{ k_j \times (q+n_\eta)}
\]
is redundant with respect to the constraint
\[
{\scriptsize\begin{bmatrix} 
M_{k_i}(\lambda_i, t_0) & M_{k_i}(\lambda_i, t_1) & \ldots & M_{k_i}(\lambda_i, t_{T-1})
\end{bmatrix}}\mathcal{G}=0_{ k_i \times (q+n_\eta)}.
\]
Similar considerations apply to the complex conjugate eigenvalues $\lambda_{r+i},  \lambda_{r+i}^\star$.
This argument shows that in $0=M\mathcal{G}$, for every real eigenvalue $\lambda_i$, we can discard all those constraints that correspond to the block-rows 
\[
\begin{bmatrix} 
M_{k_j}(\lambda_j, t_0) & M_{k_j}(\lambda_j, t_1) & \ldots & M_{k_j}(\lambda_j, t_{T-1})
\end{bmatrix}
\]
of $M$ corresponding to the eigenvalues $\lambda_j=\lambda_i$ with $k_j \le k_i$. Similarly, 
for every pair of complex conjugate eigenvalues $\lambda_{r+i},  \lambda_{r+i}^\star$, we can discard all those constraints that correspond to the block-rows 
\begin{align*}
\left[\begin{matrix} 
M_{k_{r+j}}(\lambda_{r+j},  \lambda_{r+j}^\star,t_0) & M_{k_{r+j}}(\lambda_{r+j},  \lambda_{r+j}^\star,t_1) & \ldots \end{matrix}\right.\\ 
\left.\begin{matrix} M_{k_{r+j}}(\lambda_{r+j},  \lambda_{r+j}^\star,t_{T-1})
\end{matrix}\right]
\end{align*}
of $M$ corresponding to the eigenvalues $\lambda_{r+j}=\lambda_{r+i},  \lambda_{r+j}^\star=\lambda_{r+i}^\star$, with $k_{r+j} \le k_{r+i}$. 
In doing so, we reduce the constraint 
$0=M\mathcal{G}$ to $0=\mathcal{M}\mathcal{G}$ where $\mathcal{M}$ is the matrix extracted from $M$ after removing the block-rows as indicated before. Bearing in mind that in the minimal polynomial $m_S(\lambda)$ of $S$ the degree of the monomials $s-\lambda_i$ is equal to the order of the largest Jordan block associated with the eigenvalue $\lambda_i$, we conclude that  the reduced matrix $\mathcal{M}$ has dimensions ${\color{black}\delta}\times T$, where {\color{black}$\delta$} is the degree of the minimal polynomial $m_S(\lambda)$. {\color{black} Summarizing, the above discussion leads to the following consequence of Lemma \ref{lem:dd-representation}.}

\begin{corollary}\label{cor:dd-representation}
Let the matrix  $\mathcal{M} \in \mathbb R^{d \times T}$ be obtained from $M$ in \eqref{matrix-M} by maintaining, for every distinct (i.e., counted without multiplicities) real ($\lambda_i$) and complex conjugate ($\lambda_{r+i}$, $\lambda_{r+i}^\star$) eigenvalues, the block-rows with largest indices $k_i$ and $k_{r+i}$. 
Consider any matrices $\mathcal{K} \in \mathbb R^{m \times (q+n_\eta)}$,
$\mathcal{G} \in \mathbb R^{T \times (q+n_\eta)}$ such that 
\begin{equation} \label{eq:GK-exosystem-minimal}
\begin{bmatrix} \mathcal{K} 
\\ I_{q+n_\eta} \\ 
0_{ \delta \times (q+n_\eta)}
\end{bmatrix} 
= 
\begin{bmatrix} 
U_0 \\ 
\mathcal{Z}_0 \\ 
\mathcal{M}\end{bmatrix} 
\mathcal{G}
\end{equation}
holds. Then the system \eqref{augm-sys2} under the control law \eqref{state-feedback} results in
the closed-loop dynamics \eqref{eq:GK_closed_noise_periodic}. 
\qedd
\end{corollary}

{\color{black} 
Corollary \ref{cor:dd-representation} leads to a data-based representation \eqref{eq:GK_closed_noise_periodic} of the closed-loop system, where the vector field 
$\mathcal{Z}_1 \mathcal{G} \mathcal{Z}(x, \eta)$ depends on the {\em known} matrix $\mathcal{Z}_1$, with $\mathcal{G}$ being a decision variable to be used in the design of the controller. This disturbance-free representation is remarkable because the dataset is affected by unmeasured disturbances due to the presence of $w$ acting on the plant's dynamics. Hence, the result highlights the {\em disturbance-filtering} property guaranteed by the constraint $0_{ \delta \times (q+n_\eta)}=\mathcal{M}\mathcal{G}$ in \eqref{eq:GK-exosystem-minimal}. Without that, the data-dependent closed-loop system would become 
\[
\begin{bmatrix}\dot x \\ \dot \eta  \end{bmatrix}= (\mathcal{Z}_1- \mathcal{P}W_0)\mathcal{G}\mathcal{Z}(x,\eta)+\mathcal{P}w,
\] 
where $W_0$ is the  {\it unknown} matrix defined in \eqref{eq:data4}. Besides the additional need for some a priori information on the matrix $\mathcal{P}$, the presence of $W_0$ poses extra challenges for controller design (cf.~\cite[Section V]{hu2024enforcing}), which are avoided by our disturbance filtering method.

\section{Data-based design of 
{\color{black} a harmonic} 
regulator}\label{sec:db-design-regulator}

In this section we {\color{black} show how to} design {\color{black}a} controller gain $\mathcal{K}$ {\color{black} for use in} \eqref{state-feedback} that enforces a contraction property of the closed-loop system \eqref{augm-sys2},\eqref{state-feedback} {\color{black}provided that the plant nonlinearities satisfy a certain Lipschitz continuity condition} (Proposition \ref{thm:contractivity-extended}). In this way the system  exhibits desirable properties when forced by periodic exogenous signals {\color{black}(\cite[Property 3]{pavlov2005convergent}, \cite[Theorem 5]{sontag2010contractive}, \cite[Theorem 3.8]{FB-CTDS})}, which are at the basis of the 
{\color{black} harmonic}
output regulation result formalized in Theorem \ref{cor:integr-control} below. 

{\color{black} We first define the contraction property of interest, already considered in \cite{hu2024enforcing} in the context of data-driven control.
\begin{definition}\label{def-exp-contract}
The closed-loop system \eqref{augm-sys2},\eqref{state-feedback}
\be\label{closed-loop-w-input}
\begin{bmatrix}  
\dot x \\
\dot \eta
\end{bmatrix}
= (\mathcal{A}+\mathcal{B}\mathcal{K}) \mathcal{Z}(x, \eta)
+
\mathcal{P}w,
\ee
with the input $w(t)\in \R^{n_w}$, is exponentially contractive  on 
$\mathbb{R}^n\times \mathbb{R}^{n_\eta}$ with respect to the constant positive definite metric $\mathcal{P}_1\in \mathbb{S}^{(n+n_\eta)\times (n+n_\eta)}$ if there exists $\beta>0$ such that
\[\ba{l}
\left[(\mathcal{A} +\mathcal{B} \mathcal{K}) \frac{{\partial \mathcal{Z}}}{\partial (x,\eta)}\right]^\top \!\!\!
\mathcal{P}_1^{-1}
+\mathcal{P}_1^{-1}\!\!
\left[(\mathcal{A} +\mathcal{B} \mathcal{K}) \frac{\partial \mathcal{Z}}{\partial (x,\eta)}\right]\!\!\preceq \!-\beta \mathcal{P}_1^{-1}. 
\ea\] 
for all $(x,\eta)\in \mathbb{R}^n\times \mathbb{R}^{n_\eta}$. \qedd \end{definition}
Due to the variety of contraction-related concepts, a clarification about the adopted terminology in relation with recent contributions is useful. In \cite[Proof of Theorem 1]{duvall2024global} adapted to our case, the system \eqref{closed-loop-w-input} is said to be uniformly contractive with respect to a (symmetric uniformly positive definite) metric $\mathcal{M}(x)$, uniformly over all inputs $w$ with values in $W\subset \R^{n_w}$, if there exists $\beta>0$ such that, for every input $w$ with values in $W$, all $(x,\eta)\in \mathbb{R}^n\times \mathbb{R}^{n_\eta}$ and all $t$, it holds that 
\[\ba{l}
\left[(\mathcal{A} +\mathcal{B} \mathcal{K}) \frac{{\partial \mathcal{Z}}}{\partial (x,\eta)}\right]^\top \!\!\!
\mathcal{M}(x)
+\mathcal{M}(x)\!\!
\left[(\mathcal{A} +\mathcal{B} \mathcal{K}) \frac{\partial \mathcal{Z}}{\partial (x,\eta)}\right]\!\!\\[2mm]
\hspace{4.5cm}+\dot{\mathcal{M}}(x, w(t))\preceq \!-\beta \mathcal{M}(x), 
\ea\]
where, for any pair of indices $i,j$, $\dot{\mathcal{M}}_{ij}(x, w(t))=\frac{\partial \mathcal{M}_{ij}}{\partial x}[\mathcal{A}+\mathcal{B}\mathcal{K}) \mathcal{Z}(x, \eta)
+
\mathcal{P}w(t)]$.
Specialising this notion to the case of a constant metric $\mathcal{P}_1^{-1}$, the inequality above becomes identical to the one in Definition \ref{def-exp-contract}. Because of the choice of a constant metric, the dependence on the input $w$ disappears and the property of contractivity of being uniform over all inputs $w$ is automatically satisfied. This motivates the decision to drop the ``uniformity" qualifier from the definition. We added instead the ``exponential" qualifier, because  the requirement in Definition \ref{def-exp-contract} is stronger than standard contractivity and referred to as such in e.g.~\cite[Definition 1]{aylward2008stability} (for autonomous systems). 
}

\begin{proposition} \label{thm:contractivity-extended}
Consider the closed-loop system \eqref{augm-sys2},\eqref{state-feedback}.  Assume that a matrix  $R_Q\in \mathbb R^{n \times r}$ is known 
such that
\be\label{asspt}
\frac{\partial Q}{\partial x}(x)^\top \frac{\partial Q}{\partial x}(x)\preceq R_Q R_Q^\top \qquad \text{ for all 
$x\in \mathbb{R}^n$.
}
\ee
Set $\mathcal{R}_Q:=\left[\begin{smallmatrix} R_Q \\ 0_{ n_\eta\times r}\end{smallmatrix}\right]$.
Consider 
the following SDP in the decision variables $0\prec \mathcal{P}_1 \in \mathbb S^{(n+n_\eta) \times (n+n_\eta)}$,
$\mathcal{Y}_1 \in \mathbb R^{T \times (n+n_\eta)}$, 
$\mathcal{G}_2 \in \mathbb R^{T \times (q-n)}$, $\alpha \in \mathbb R_{> 0}$: 
\begin{subequations}
\label{eq:SDP-extended}
\begin{alignat}{6}
& \mathcal{Z}_0 \mathcal{Y}_1 = \begin{bmatrix} \mathcal{P}_1 \\ 0_{(q-n) \times (n+n_\eta)} \end{bmatrix} \,,
\label{eq:SDP1-extended} \\
& 
\begin{bmatrix} 
\mathcal{Z}_1 \mathcal{Y}_1 +(\mathcal{Z}_1 \mathcal{Y}_1)^\top + 
\alpha I_{ n+n_\eta }
& \mathcal{Z}_1 \mathcal{G}_2 & \mathcal{P}_1  \mathcal{R}_Q  \\[1mm] 
(\mathcal{Z}_1 \mathcal{G}_2)^\top  & - I_{ q-n } & 0_{ (q-n) \times r}  \\
(\mathcal{P}_1\mathcal{R}_Q )^\top & 0_{r \times  (q-n) } & - I_{r} 
\end{bmatrix}  \preceq 0 
\,, \label{eq:SDP2-extended} \\[1mm] 
& \mathcal{Z}_0 \mathcal{G}_2 = \begin{bmatrix} 0_{ (n+n_\eta) \times (q-n) } \\ I_{ q-n } \end{bmatrix},
\label{eq:SDP4-extended}\\
& 
\mathcal{M} 
\begin{bmatrix} 
\mathcal{Y}_1 & 
\mathcal{G}_2 
\end{bmatrix}  
= 
0_{ d \times (q+n_\eta)}\,. \label{eq:SDP5-extended} 
\end{alignat}
\end{subequations} 
If the program is feasible then \eqref{state-feedback}
with $\mathcal{K}$ defined as 
\begin{eqnarray} \label{eq:K_SDP-extended}
\mathcal{K}= U_0 \begin{bmatrix} \mathcal{Y}_1 \mathcal{P}_1^{-1}
& \mathcal{G}_2 \end{bmatrix}
\end{eqnarray}
renders the closed-loop dynamics \eqref{augm-sys2},\eqref{state-feedback} exponentially contractive on 
$\mathbb{R}^n\times \mathbb{R}^{n_\eta}$. 
\qedd
\end{proposition} 

{\it Proof.} It can be proven similarly to  \cite[Theorem 4]{hu2024enforcing} and the proof is therefore omitted. \qedp

\smallskip

Given a compact convex set $\mathcal{X}\subset \R^n$, a local version of the result above, which holds on $\mathcal{X}\times  \R^{n_\eta}$, can also be shown. In this case, thanks to the  continuity of the functions in $\partial Q/\partial x$, a matrix $R_Q$ that satisfies the condition \eqref{asspt} always exists. Below, for the sake of simplicity, we only focus on a global result.

\smallskip

The result below provides a solution to the  data-driven 
{\color{black} harmonic}
output regulation problem. 
\begin{theorem}\label{cor:integr-control}
Let the conditions of {\color{black}Proposition} \ref{thm:contractivity-extended} hold. 
Let Assumption \ref{assum:w} hold. 
Then the control law \eqref{state-feedback}
with $\mathcal{K}$ defined as in \eqref{eq:K_SDP-extended}
is such that  the solutions of the closed-loop dynamics  
\eqref{augm-sys2},\eqref{state-feedback}
are bounded for all $t\ge 0$ and the regulation error $e(t)$ converges exponentially to a bounded $D$-periodic signal $e_*(t)$, {\color{black}whose components have each the first $\ell+1$ coefficients of their Fourier series equal to zero.} 
Moreover, for any initial condition $w(0)$ of \eqref{plant1-nl}, {\color{black} for every $k=1,2,\ldots, p$,} there {\color{black}exist constants} $\bar e_{{\color{black}*k}}$ such that 
\be\label{l2-gain}
\displaystyle 
\frac{1}{D}
\int_0^{D} e_{{\color{black}*k}}(t)^2 dt 
\le \frac{ \bar e_{{\color{black}*k}}^2}{\left(\frac{2\pi}{D}\right)^2  (\ell+1)^2}.
\ee
{\color{black}where $e_{{\color{black}*k}}$ is the component $k$ of $e_*$.}
\end{theorem} 

{\it Proof.} By Proposition \ref{thm:contractivity-extended}, the closed-loop system \eqref{augm-sys2},\eqref{state-feedback},  namely
\be\label{system-(9b)-(16)}
\begin{bmatrix}  
\dot x \\
\dot \eta
\end{bmatrix}
=
(\mathcal{A} +\mathcal{B} \mathcal{K})
\mathcal{Z}(x, \eta)+
\mathcal{P} w,
\ee
is exponentially contractive on $\mathbb{R}^n\times \mathbb{R}^{n_\eta}$. The system is driven by the input $w$, which exists and is bounded for all 
{\color{black}$t\in \R$,}
and $D$-periodic, by Assumption \ref{assum:w}. 
The existence and boundedness of $w(t)$ for all {\color{black}$t\in \R$}
imply that, for a fixed value of $(x,\eta)$, the right-hand side of 
\eqref{system-(9b)-(16)}
is bounded for all 
{\color{black}$t\in \R$,}
as required by \cite[Theorem 1]{pavlov2004convergent} (see also \cite[Theorem 6]{hu2024enforcing}). Then there exists a unique solution $(x_*(t),  \eta_*(t))$ defined and bounded for all 
{\color{black}$t\in \R$, dependent on $w$, }
which is globally 
 exponentially stable and $D$-periodic {\color{black}(\ref{AppB})}. By definition, 
$\eta_*(t)$ satisfies $\dot{\eta}_*(t)=\Phi {\eta}_*(t) + G e_*(t)$, where $e_*(t)=C_e Z(x_*(t))+Q_e w(t)$. By $D$-periodicity of  $x_*(t), w(t)$ and the continuity of $Z(x)$,  $e_*(t)$ is $D$-periodic. By Lemma \ref{zero-fourier-coefficients}(i), (ii) in the Appendix (in particular, bearing in mind Eq.~\eqref{sub-im})
and the definition of $\Phi,G$, the first $\ell+1$ coefficients of the Fourier series of each component of $e_*(t)$ must be zero. Furthermore,   $\dot e_*(t)=C_e\frac{\partial Z}{\partial x}(x_*(t)) (A Z(x_*(t))+B \mathcal{K} \mathcal{Z}(x_*(t), \eta_*(t)))+Q_e S w(t)$. Thanks to the boundedness  and $D$-periodicity of $x_*(t), \eta_*(t), w(t)$, and the continuous differentiability of $Z(x)$, 
$\dot e_*(t)$ is $D$-periodic and {\color{black} for each $k=1,2,\ldots, p$} there exists $\bar e_{{\color{black}*k}}$ such that {\color{black}$\max_{t} |\dot e_{{\color{black}*k}}(t)|=\bar e_{{\color{black}*k}}$}.  By Lemma \ref{zero-fourier-coefficients}(iii),  the inequality \eqref{l2-gain} holds. 
\qedp

\smallskip

{\color{black}
We underscore two aspects of Theorem \ref{cor:integr-control}:

\smallskip 

\noindent (i) The inequality \eqref{l2-gain} does not imply that the $\mathcal{L}_2$-norm of the components of $e_*$ can be made arbitrarily small by increasing the number $\ell$ of oscillators included in the internal model. In fact, increasing $\ell$ leads to a different closed-loop dynamics  
\eqref{augm-sys2},\eqref{state-feedback}, hence to a different steady state solution $(x_*(t),  \eta_*(t))$, which in general results in a new bound $\bar e_{{\color{black}*k}}$. Nonetheless, as noted in \cite[p.~2]{astolfi2022harmonic}, guaranteeing the first coefficients of the Fourier series of $e_*$ to be zero can be sufficient in practice.

\smallskip

\noindent (ii) The solution of the harmonic regulation problem based on data originates from the solution of the SDP \eqref{eq:SDP-extended}. This is a different SDP than the one required to design data-driven stabilizers or 
controllers that enforce contractivity, in that the constraint \eqref{eq:SDP5-extended} is added. Such a constraint, suggested by Corollary \ref{cor:dd-representation}, is the key to have a controller that is independent of the amplitude and phase of the disturbance and takes into account the specifics of the output regulation problem, where the plant dynamics are affected by a disturbance generated by a known exosystem.  

}

\begin{example}\label{ex_nonlinear}
Consider the dynamics of a one-link robot arm \cite[Section 4.10]{isidori1995nonlinear}
\begin{subequations}\label{robot.arm}   
\begin{align}
&\dot x_1 = x_2 +d_1
\\
&\dot x_2 = -\frac{K_c}{J_2}x_1 - \frac{F_2}{J_2}x_2 +\frac{K_c}{J_2 N_c}x_3 - \frac{mgd}{J_2}\cos x_1 +d_2
\\
&\dot x_3 = x_4 + d_3
\\
&\dot x_4 = -\frac{K_c}{J_1 N_c}x_1 + \frac{K_c}{J_1 N_c^2}x_3 - \frac{F_1}{J_1}x_4 + \frac{1}{J_1}u + d_4
\\
& e = x_1 - r.
\end{align}
\end{subequations}
where $x_1$, $x_3$ denote the angular positions of the link and of the actuator shaft, respectively,  $u$ is the torque produced at the actuator axis and $e$ is the error variable. $d:=[d_1\; d_2\; d_3\; d_4]^\top$ is a vector of disturbance signals. The control objective is to regulate $x_1$ to a desired reference signal $r$. Below we define $d,r$ as functions of the state $w$ of the exosystem. We assume that  $r(t) = \cos(t)$ and  the system is corrupted by the periodic disturbance $d(t)= \begin{bmatrix} 0.2 & \sin(t)  & \cos(2t) & 0.5+3\sin(t + \frac{\pi}{3}) \end{bmatrix}^\top$.  Let $w_1(t) = 1$, $w_2(t) = \sin(t)$, $w_3(t) = \cos(t)$, $w_4(t) = \sin(2t)$ and $w_5(t) = \cos(2t)$, and then we have
\[\ba{rl}
d(t)&={\scriptsize\begin{bmatrix}
   0.2& 0 & 0 &0 & 0\\
   0& 1 & 0 &0 & 0\\
   0& 0 & 0 &0 & 1\\
   0.5& \frac{3}{2} & \frac{3\sqrt{3}}{2} &0 & 0\\
\end{bmatrix}} w(t) =: Pw(t)
\\
r(t)&={\scriptsize\begin{bmatrix}
   0& 0 & 1 &0 & 0
\end{bmatrix}} w(t) =: Q_e w(t),
\ea
\]
and $w \in \mathbb{R}^5$ satisfies
\[
\dot w = {\scriptsize\begin{bmatrix} 
0& 0& 0& 0& 0\\
0& 0& 1& 0& 0 \\
0& -1& 0& 0& 0\\
0& 0& 0& 0 & 2 \\
0& 0& 0& -2 & 0 \end{bmatrix}} w =: Sw.
\]
Note that the solution $w(t)$ of the exosystem $\dot w = Sw$ is a bounded and periodic function with period $T=2\pi$, and hence the fundamental frequency in the internal model \eqref{im-k} is set to 1 \text{Hz}. Besides, without loss of generality, we set $\gamma=1$ and $N = \begin{bmatrix}
  0 & 1
\end{bmatrix}^\top$ in the internal model \eqref{im-k}. 
We introduce $Q(x)=  \cos x_1 $. If we set $\mathcal{X} = \mathbb{R}^4$, then \eqref{asspt} is satisfied with $R_Q = R_Q ^\top =
\smat{
1& 0& 0& 0\\
0& 0& 0& 0  \\
0& 0& 0& 0\\
0& 0& 0& 0}$. 

We  collect data from the system setting the parameters to $K_c=0.4$, $F_2=0.15$, $J_2=0.2$, $N_c=2$, $F_1=0.1$, $J_1=0.15$, $m=0.4$, $g=9.8$ and $d=0.1$. Here, we consider different scenarios by varying the number of oscillators $\ell$ from $0$ to $4$. For each scenario, we collect $T =40$ samples, i.e. $\mathbb D := \left\{ (x_i, u_i, e_i, \dot x_i, \eta_i )\right\}_{i=0}^{39}$,  by running an experiment with input uniformly distributed in  $[-0.1,0.1]$,
and with an initial state within the same interval. The  matrix $\mathcal{M}$ is given as 
\[
\mathcal{M} = 
\begin{bmatrix}
1 &  1 & \ldots &  1\\ 
\cos(t_0)&  \cos(t_1) & \ldots & \cos(t_{39})\\ 
\sin(t_0)&  \sin(t_1) & \ldots & \sin(t_{39})\\
\cos(2t_0)&  \cos(2t_1) & \ldots & \cos(2t_{39})\\ 
\sin(2t_0)&  \sin(2t_1) & \ldots & \sin(2t_{39})
\end{bmatrix}
.
\] 
The SDP \eqref{eq:SDP-extended} is feasible and under the obtained controller $\mathcal{K}$, the asymptotic value of the error variable $e$ is shown in Table \ref{tab1}.
Note that as the number of oscillators increases, the 
regulation objective is improved. 

\begin{figure*}
\begin{equation}
\label{K}
\mathcal{K} = \smat{ -61.4992&  -58.6366&  -38.5176 &  -5.6908 & -16.2628 &  13.9920 & -11.3739 &  22.8700 &  11.4681 &   1.7275  &  4.5851  &  1.2868  &  1.9170  &  1.9648 }
\hrulefill
\end{equation}
\end{figure*}

\begin{table}[!ht]
\centering
\caption{Asymptotic value of the 
error $e$
depending on the number of oscillators $\ell$. }\label{tab1}
$\begin{array}{lcc}
\hline \ell & \text{lim \!sup}_{t \to +\infty} |e(t)| \\
\hline 0 &  1.385\\
\hline 1 & 0.210\\
\hline 2 & 8.8 \times 10^{-4}\\
\hline 3& 8.1 \times 10^{-4}\\
\hline 4& 5.7 \times 10^{-6}\\
\hline
\end{array}
$
\end{table}

\begin{figure}[ht!] \centering
\includegraphics[scale=0.5]{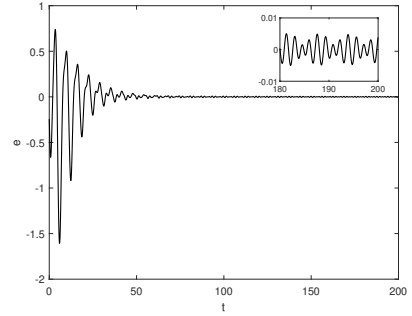}
   \caption{Evolution of the error $e$ with an internal model composed of 4 oscillators. }\label{nonlinear}
\end{figure}
\end{example}

\smallskip

Semidefinite programs to design convergence-based controllers that solve the global output regulation problem have been also proposed in the context of model-based solutions  \cite{pavlov2007global}.  The solution that we propose in Proposition \ref{thm:contractivity-extended} and Theorem \ref{cor:integr-control} appears to be substantially different from this previous work.

{\color{black}
\subsection{The regulator design under a relaxed Assumption \ref{assum:w}}\label{subsec:rel-Asspt1}
Here we consider the case in which we differentiate between an unmeasured disturbance $d$ and a measured reference signal $r$ in $w$, i.e.,  
$w=\left[d^\top \; r^\top \right]^\top$. We assume that $d$ is generated by a linear exosystem,  say $\dot d=S_d d$, where $S_d$ is such that the solution $d(t)$ satisfies Assumption \ref{assum:w}, while $r(t)$ exists for all 
{\color{black} $t\in \R$},
is bounded, $D$-periodic and can be measured, but need not be generated by a finite-dimensional linear exosystem. We then rewrite the plant's dynamics $\dot x = AZ(x) +Bu +Pw$ as 
\[
\dot x = AZ(x) +Bu +P_d d+P_r r
\] 
and the error's equations $e=C_e Z(x) + Q_{e} w$ as 
\[
e=C_e Z(x) + Q_{ed} d + Q_{er} r.
\] 
As a result, the matrix $\mathcal{P}$ in Eq.~\eqref{augm-sys2} can be split as 
\[
\mathcal{P}=
\begin{bmatrix}  
P_d & P_r \\
GQ_{ed} & GQ_{er}
\end{bmatrix}=:
\begin{bmatrix}  
\mathcal{P}_d & \mathcal{P}_r 
\end{bmatrix}.
\] 
When we run the offline experiment to collect data, the matrix $W_0$ in \eqref{eq:data4} can be written as
\[
 W_0 = \begin{bmatrix} d (t_0) &\ldots & d(t_{T-1}) \\
                                          r (t_0) &\ldots & r(t_{T-1})
                                          \end{bmatrix}=:\begin{bmatrix} D_0 \\ R_0\end{bmatrix},
\]
where $D_0$ is an unknown matrix while $R_0$ is a known one. Then the relation \eqref{data-matrices-identity} becomes
\be\label{data-matrices-identity-dist-ref}
\mathcal{Z}_{1} = \mathcal{A}  \mathcal{Z}_{0} +\mathcal{B} U_0 +\mathcal{P}_d D_0+\mathcal{P}_r R_0\,.
\ee
Lemma \ref{lem:forma-nuova-risposta-esosistema} applied to $\dot d = S_d d$ leads to the partition $D_0=
\mathcal{T}_d L_d M_d$, where the matrices appearing in the factorization have a similar meaning as those in \eqref{W0-identity}. Hence, from \eqref{data-matrices-identity-dist-ref} we have
\be\label{data-matrices-identity-dist-ref-2}
\mathcal{Z}_{1} = \mathcal{A}  \mathcal{Z}_{0} +\mathcal{B} U_0 +\mathcal{P}_d \mathcal{T}_d L_d  M_d +\mathcal{P}_r R_0\,.
\ee
Using this identity, and extracting the matrix $\mathcal{M}_d$ from $M_d$ with the arguments preceding Corollary \ref{cor:dd-representation},  we can show that Corollary \ref{cor:dd-representation} holds under the following modified version of the condition \eqref{eq:GK-exosystem-minimal}:
\be\label{eq:GK-exosystem-minimal-relax_Asspt1}
\begin{bmatrix} \mathcal{K} 
\\ 
I_{q+n_\eta} 
\\ 
0_{{\delta_d} \times (q+n_\eta)}
\\
0_{n_r \times (q+n_\eta)}
\end{bmatrix} 
= 
\begin{bmatrix} 
U_0 \\ 
\mathcal{Z}_0 \\ 
\mathcal{M}_d\\
R_0
\end{bmatrix} 
\mathcal{G}
\ee
where ${\delta_d}$ is the degree of the minimal polynomial of $S_d$ and $n_r$ is the dimension of the vector $r$. This implies that if we replace \eqref{eq:SDP5-extended} with 
\[
\begin{bmatrix}
\mathcal{M}_d\\
R_0 
\end{bmatrix}
\begin{bmatrix} 
\mathcal{Y}_1 & 
\mathcal{G}_2 
\end{bmatrix}  
= 
0_{(\delta_d+n_r) \times (q+n_\eta)}
\]
Theorem \ref{cor:integr-control} would still hold, this time under a relaxed Assumption \ref{assum:w}, which only requires $d$ to be generated by a linear exosystem and have the properties stated in Assumption \ref{assum:w}. 

On the other hand, a drawback of the new condition \eqref{eq:GK-exosystem-minimal-relax_Asspt1} is that it depends on the specific reference signal that is used during the data acquisition phase. Hence, if a different reference must be tracked during the control execution phase, a new dataset must be acquired. 
Finally, we observe that if one is solving a tracking problem for which no disturbance is present and the way the reference enters the dynamics is known (i.e.,  $\mathcal{P}_r$ is known), then one can draw the same conclusions of Theorem 1 by solving \eqref{eq:SDP-extended} where the matrix $\mathcal{Z}_1$ is replaced by $\mathcal{Z}_1-\mathcal{P}_r R_0$ and without the condition \eqref{eq:SDP5-extended}. 
}

\section{
The case of linear systems}\label{sec:linear-systems}
We specialize the previous result to the case of linear systems. We do this for two reasons. First, in this case the previous arguments lead to the solution of the {\color{black} {\em exact global output regulation} problem.}
Second, the analysis provides insights into the SDP \eqref{eq:SDP-extended} derived for the 
{\color{black} nonlinear} output regulation result, and appears to be an improvement over existing results.

In the case of linear systems, the controlled plant is 
\begin{subequations}
\label{plant}
\begin{alignat}{3}
\dot w=&Sw \label{plant1}
\\
\dot x =& Ax +Bu +Pw\label{plant2}
\\
\begin{bmatrix}  
e\\
x\\
\eta
\end{bmatrix} 
=
& 
\begin{bmatrix}  
\begin{bmatrix}
C_e \\ I_n
\end{bmatrix} 
& 0\\
0 & I
\end{bmatrix}
\begin{bmatrix}  
x \\
\eta
\end{bmatrix}
+
\begin{bmatrix}  
\begin{bmatrix}
Q_e \\ 0
\end{bmatrix} \\
0
\end{bmatrix}w\label{plant3}
\end{alignat}
\end{subequations}
and the regulator is 
\be\label{lin-outp-regulator}
\ba{rl}
\dot \eta =& \Phi\eta+Ge\\
u =& \mathcal{K}\begin{bmatrix}
x\\ \eta
\end{bmatrix}.
\ea
\ee
{\color{black} The objective is to find $\mathcal{K}$ in} \eqref{lin-outp-regulator} such that  (i) the autonomous system obtained from \eqref{plant2}, \eqref{lin-outp-regulator} setting $w=0$ is asymptotically stable  and (ii) $e(t)\to 0$ as $t\to +\infty$. {\color{black} As is traditional in the linear case,} the matrices $\Phi, G$ defining the internal model are different from before. {\color{black} In particular, denoting} by $m_S(\lambda)= s_0+s_1 \lambda+\ldots+s_{d-1} \lambda^{d-1}+\lambda^{d}$ the minimal polynomial of $S$, the matrices $\Phi, G$ take  the form \cite[Eq.~(4.23)-(4.24)]{isidori2017lectures}
\be\label{im-matrices}
\Phi=
\begingroup 
\setlength\arraycolsep{1.5pt}
\begin{bmatrix}  
0 & I_p & 0 & \ldots & 0\\
0 & 0 & I_p & \ldots & 0\\
\vdots & \vdots & \vdots & \ddots & \vdots\\
0 & 0 & 0 & \ldots & I_p\\
-s_0 I_p &-s_1 I_p &-s_2 I_p &\ldots & -s_{d-1} I_p 
\end{bmatrix},
\endgroup 
G = 
\begin{bmatrix}  
0 \\ 0\\ \vdots \\ 0 \\I_p
\end{bmatrix}.
\ee
Note that $\Phi\in R^{n_\eta\times n_\eta}$, where in this case $n_\eta=pd$.
We collect data 
\begin{equation} \label{dataset-extended}
\mathbb D := \left\{ (x_i,u_i, e_i, \dot x_i, \eta_i)\right\}_{i=0}^{T-1}
\end{equation} 
from systems \eqref{plant2}, \eqref{plant3}, \eqref{lin-outp-regulator} through an experiment and, by a slight abuse of notation,  define the matrix
\[
\mathcal{Z}_0:= \begin{bmatrix}
x(t_0) &  x(t_1) & \ldots &  x(t_{T-1})\\ 
\eta(t_0)&  \eta(t_1) & \ldots & \eta(t_{T-1})
\end{bmatrix}
\]
in addition to the data matrices $U_0$ and $\mathcal{Z}_1$ defined in \eqref{eq:data}. 
These matrices satisfy the identity \eqref{data-matrices-identity}, where $\mathcal{B}, \mathcal{P}$ are as in \eqref{augm-sys2}, whereas $\mathcal{A}=\left[\begin{smallmatrix} A & 0_{n\times n_\eta}\\ G C_e &\Phi  \end{smallmatrix}\right]$. A similar result as Corollary \ref{cor:dd-representation} holds, namely, for any matrices $\mathcal{K} \in \mathbb R^{m \times (n+n_\eta)}$,
$\mathcal{G} \in \mathbb R^{T \times (n+n_\eta)}$ such that 
\begin{equation} \label{eq:GK-exosystem-minimal-linear}
\begin{bmatrix} \mathcal{K} 
\\ I_{n+n_\eta} \\ 
0_{ d \times (n+n_\eta)}
\end{bmatrix} 
= 
\begin{bmatrix} 
U_0 \\ 
\mathcal{Z}_0 \\ 
\mathcal{M}\end{bmatrix} 
\mathcal{G}
\end{equation}
holds,  the system 
 \begin{subequations}
\label{augm-sys-lin}
\begin{alignat}{2}
\begin{bmatrix}  
\dot x \\
\dot \eta
\end{bmatrix}
=& 
\begin{bmatrix}
A & 0_{n\times n_\eta}\\
\,G C_e & \Phi 
\end{bmatrix}
\begin{bmatrix}  
x \\
\eta
\end{bmatrix}
+
\begin{bmatrix}  
B \\
0
\end{bmatrix}
u
+
\begin{bmatrix}  
P \\
GQ_e
\end{bmatrix}
w\label{augm-sys-lin1}
\\
=&\mathcal{A} 
\begin{bmatrix}  
x \\
\eta
\end{bmatrix}
+
 \mathcal{B} u
 +
\mathcal{P} w \label{augm-sys-lin2}
\end{alignat}
\end{subequations}
with $u = \mathcal{K}\left[\begin{smallmatrix}
x\\ \eta
\end{smallmatrix}\right]$ results in
the closed-loop dynamics 
\be\label{eq:GK_closed_noise_periodic-linear}
\begin{bmatrix}  
\dot x \\
\dot \eta
\end{bmatrix}
= 
\mathcal{Z}_1 \mathcal{G} 
\begin{bmatrix}  
x \\
\eta
\end{bmatrix}
+\mathcal{P} w.
\ee
We have the following result:
\begin{theorem} \label{thm:contractivity-extended-linear}
Consider the controlled plant \eqref{plant}
and  
the following SDP in the decision variables $\mathcal{X}  \in \mathbb S^{(n+n_\eta) \times (n+n_\eta)}$,
$\mathcal{Y} \in \mathbb R^{T \times (n+n_\eta)}$: 
\begin{subequations}
\label{eq:SDP-extended-linear}
\begin{alignat}{6}
&\mathcal{X}\succ 0\label{eq:SDP0-extended-linear}\,, \\
&
\begin{bmatrix} 
\mathcal{X}
\\ 0_{ d \times (n+n_\eta)}\end{bmatrix} = 
\begin{bmatrix} 
\mathcal{Z}_0 \\ \mathcal{M}\end{bmatrix} \mathcal{Y} \,,
\label{eq:SDP1-extended-linear}
\\
& 
\mathcal{Z}_1 \mathcal{Y}+ \mathcal{Y}^\top \mathcal{Z}_1^\top
\prec 0
. \label{eq:SDP2-extended-linear}
\end{alignat}
\end{subequations} 
If the program is feasible then the control law \eqref{lin-outp-regulator} with the matrices $\Phi, G$ as in \eqref{im-matrices} and
 $\mathcal{K}$ defined as 
\begin{eqnarray} \label{eq:K_SDP-extended-linear}
\mathcal{K}= U_0  \mathcal{Y}  \mathcal{X}^{-1}
\end{eqnarray}
solves the output regulation problem. 

Conversely, 
{\color{black}assume that} 
$\left[\begin{smallmatrix} U_0 \\ \mathcal{Z}_0 \\ \mathcal{M}\end{smallmatrix}\right]$ {\color{black}has} full row rank.  If the output regulation problem is solvable, then there exist $\mathcal{X}  \in \mathbb S^{(n+n_\eta) \times (n+n_\eta)}$ and 
$\mathcal{Y} \in \mathbb R^{T \times (n+n_\eta)}$ that solve the SDP \eqref{eq:SDP-extended-linear}. 
\end{theorem} 

{\em Proof.} Set $\mathcal{G}:=\mathcal{Y}\mathcal{X}^{-1}$. Then, by \eqref{eq:SDP1-extended-linear}, \eqref{eq:K_SDP-extended-linear}, the condition \eqref{eq:GK-exosystem-minimal-linear} is satisfied. Hence,  
bearing in mind \eqref{eq:GK_closed_noise_periodic-linear},
the dynamics \eqref{plant1}, \eqref{plant2}, \eqref{lin-outp-regulator} are given by 
\be\label{autonomous-closed-loop}
\ba{rl}
\begin{bmatrix}
\dot w\\
\begin{bmatrix}
\dot x\\
\dot \eta
\end{bmatrix}
\end{bmatrix}
=&
\underbrace{
\begin{bmatrix}
S & 0 \\
\mathcal{P}&
\mathcal{Z}_1 \mathcal{G}
\end{bmatrix}
}_{=:\mathcal{A}_{cl}}
\begin{bmatrix}
w\\
\begin{bmatrix}
x\\
\eta
\end{bmatrix}
\end{bmatrix}.
\ea
\ee
{\color{black}Replacing $\mathcal{Y}$ with $\mathcal{G}\mathcal{X}$ in \eqref{eq:SDP2-extended-linear} yields that} 
$\mathcal{Z}_1 \mathcal{G}$ is Hurwitz, showing that the {\color{black}objective} (i) of the exact output regulation problem is satisfied. Since 
$\mathcal{Z}_1 \mathcal{G}=\left[\begin{smallmatrix} A+BK_x & B K_\eta\\ G C_e & \Phi\end{smallmatrix}\right]$, 
where  $\mathcal{K}= \left[\begin{smallmatrix} K_x & K_\eta\end{smallmatrix}\right]$ and $K_x\in \R^{m\times n_\eta}, K_\eta\in \R^{m\times n_\eta}$,
the rest of the proof proceeds along the thread of the model-based result. 
By Assumption \ref{assum:w}, all the eigenvalues of $S$ have zero real part, hence the system \eqref{autonomous-closed-loop} has a stable and a center eigenspace \cite[p.~100]{isidori2017lectures}. Specifically, the center eigenspace $\mathcal{V}_c$ is given by
\[
\mathcal{V}_c={\rm Im} \begin{bmatrix}
I_{n_w}\\
\Pi_x\\
\Pi_\eta
\end{bmatrix}
\]
for some matrices $\Pi_x\in \R^{n\times n_w}, \Pi_\eta\in \R^{n_\eta\times n_w}$. As $\mathcal{V}_c$ is $\mathcal{A}_{cl}$-invariant, the invariance property implies
\[
\begin{bmatrix}
S & 0 \\
\mathcal{P}
&
\mathcal{Z}_1 \mathcal{G}
\end{bmatrix}
\begin{bmatrix}
I_{n_w}\\
\Pi_x\\
\Pi_\eta
\end{bmatrix}
=\begin{bmatrix}
I_{n_w}\\
\Pi_x\\
\Pi_\eta
\end{bmatrix}
Y
\]
for some matrix $Y$. This identity is equivalent to the Sylvester equation
\be\label{sylv.eq}
\mathcal{P}+ \mathcal{Z}_1 \mathcal{G} 
\begin{bmatrix}
\Pi_x\\
\Pi_\eta
\end{bmatrix}
=
\begin{bmatrix}
\Pi_x\\
\Pi_\eta
\end{bmatrix}
S.
\ee
By a property of the Sylvester equation  
\cite[Corollary A.1.1]{knobloch1993topics}, there exist unique matrices $\Pi_x$ and $\Pi_\eta$ such that \eqref{sylv.eq} holds. 
Recall that 
$\mathcal{P}= 
\left[\begin{smallmatrix} 
P \\
G Q_e
\end{smallmatrix}\right]
$ 
and partition  $\mathcal{Z}_1 \mathcal{G}$
accordingly as 
\[
\mathcal{Z}_1 \mathcal{G}= 
\begin{bmatrix}
X_1 \\
N_1
\end{bmatrix}
\begin{bmatrix}
\mathcal{G}_1 &
\mathcal{G}_2
\end{bmatrix}
\]
where  $X_1 \in \R^{n\times T}$, $N_1 \in \R^{n_\eta\times T}$, $\mathcal{G}_1 \in \R^{T\times n}$ and $\mathcal{G}_2 \in \R^{T \times n_\eta}$.
Hence, the identity \eqref{sylv.eq} can be written as 
\[\ba{rl}
P+  X_1 \mathcal{G}_1 \Pi_x + X_1 \mathcal{G}_2  \Pi_\eta=&\Pi_x S 
\\
G Q_e+ N_1 \mathcal{G}_1  \Pi_x +  N_1 \mathcal{G}_2 \Pi_\eta = &\Pi_\eta S.
\ea
\]
{\color{black} Bearing in mind that $\mathcal{Z}_1 \mathcal{G}=\left[\begin{smallmatrix} A+BK_x & B K_\eta\\ G C_e & \Phi\end{smallmatrix}\right]=\left[\begin{smallmatrix}
X_1 \\
N_1
\end{smallmatrix}\right] 
\left[\begin{smallmatrix}
\mathcal{G}_1 &
\mathcal{G}_2
\end{smallmatrix}\right]
$, the second of the identities above becomes  $G(Q_e+C_e \Pi_x)+\Phi \Pi_\eta=\Pi_\eta S$, which} implies $C_e\Pi_x +Q_e=0$ by \cite[Lemma 4.3]{isidori2017lectures}.  Introduce the variables $\tilde x=x-\Pi_x w$, $\tilde \eta=\eta-\Pi_\eta w$, which evolve as 
\[
\begin{bmatrix}
\dot{\tilde x} \\
\dot{\tilde \eta} 
\end{bmatrix}
= 
\mathcal{Z}_1 \mathcal{G} \begin{bmatrix}
 \tilde x \\
 \tilde \eta
\end{bmatrix}, \quad 
e =  \begin{bmatrix}
C_e & 0
\end{bmatrix} \begin{bmatrix}
 \tilde x \\
 \tilde \eta
\end{bmatrix}.
\]
This shows that {\color{black}objective} (ii) of the exact global output regulation problem is satisfied and ends the proof of the first part of the statement. 

Conversely, if the output regulation problem is solvable, then by \cite[Proposition 4.2 and Lemma 4.1]{isidori2017lectures} the triplet $\left(A,B, C=\left[\begin{smallmatrix}
C_e \\ I_n
\end{smallmatrix} \right]\right)$ is stabilizable and detectable\footnote{In this case the detectability property trivially holds, since the measurements include the full state $x$.} and the nonresonance condition ${\rm rank}\left[\begin{smallmatrix} A-\lambda I & B \\ C_e & 0\end{smallmatrix}\right]=n+p$ for all $\lambda\in {\rm spectrum}(S)$  holds. By \cite[Lemma 4.2]{isidori2017lectures} these properties imply stabilizability and detectability of the augmented system 
\eqref{augm-sys-lin}
with $w=0$
 \begin{subequations}
\label{augm-sys-w=0}
\begin{alignat}{2}
\begin{bmatrix}  
\dot x \\
\dot \eta
\end{bmatrix}
=& 
\mathcal{A}\begin{bmatrix}  
x \\
\eta
\end{bmatrix}
+
\mathcal{B}u
\label{augm-sys1-w=0}
\\
y_a =: & 
\begin{bmatrix}  
\begin{bmatrix}
C_e \\ I_n
\end{bmatrix} 
& 0\\
0 & I
\end{bmatrix}
\begin{bmatrix}  
x \\
\eta
\end{bmatrix}.
\label{augm-sys3-w=0}
\end{alignat}
\end{subequations}
In particular, \eqref{augm-sys1-w=0} must be stabilizable. Let $\mathcal{K}$ be such that  
$\mathcal{A}+\mathcal{B}\mathcal{K}$ is Hurwitz. Let $\mathcal{G}$ be a matrix such that 
$\left[\begin{smallmatrix} \mathcal{K} \\ I_{n+n_\eta} \\ 0_{d\times (n+n_\eta)}\end{smallmatrix}\right]=\left[\begin{smallmatrix} U_0 \\ \mathcal{Z}_0 \\ \mathcal{M}\end{smallmatrix}\right]\mathcal{G}$. Such a matrix exists because  $\left[\begin{smallmatrix} U_0 \\ \mathcal{Z}_0 \\ \mathcal{M}\end{smallmatrix}\right]$ has full column rank. Then by Corollary \ref{cor:dd-representation} $\mathcal{A}+\mathcal{B}\mathcal{K}=\mathcal{Z}_1\mathcal{G}$, which is equivalent to the existence of $\mathcal{X}\succ 0$ such that  $\mathcal{Z}_1\mathcal{G}\mathcal{X}+\mathcal{X}\mathcal{G}^\top \mathcal{Z}_1^\top\prec 0$. Setting $\mathcal{Y}=\mathcal{G}\mathcal{X}$ shows the fulfilment of \eqref{eq:SDP2-extended-linear}. The constraint \eqref{eq:SDP1-extended-linear} holds because of $\left[\begin{smallmatrix} \mathcal{K} \\ I_{n+n_\eta} \\ 0_{d\times (n+n_\eta)}\end{smallmatrix}\right]=\left[\begin{smallmatrix} U_0 \\ \mathcal{Z}_0 \\ \mathcal{M}\end{smallmatrix}\right]\mathcal{G}$. This ends the proof.
\qedp

\smallskip

Thanks to the noise-filtering design of the matrix $\mathcal{M}$ revealed by Corollary \ref{cor:dd-representation}, the sufficient condition \eqref{eq:SDP-extended} does not present any element depending on the measurements of the exogeneous signal $w$, which is an improvement over the 
existing
results in linear output regulation theory (cf.~\cite[Eq.~(19)]{zhu2024data}). {\color{black}In fact, in one of their main results \cite[Theorem 3.2]{zhu2024data}, the design of the regulator depends on the condition \cite[Eq.~(26)]{zhu2024data}, where a matrix $\mathcal{S}$ defined in \cite[Eq.~(19)]{zhu2024data} appears. In turn, such a matrix   $\mathcal{S}$ depends on the matrix $V_-$, which contains the samples of the disturbance generated by the exosystem and affecting the plant. This is in stark contrast with our results, which do not require the measurement of the disturbance $w$, neither during the offline data collection phase, nor during the execution of the control task.} On the other hand, \cite[Theorem 3.2]{zhu2024data} seems to suggest that for the necessity part, one might renounce to  the full row rank condition of  $\left[\begin{smallmatrix} U_0 \\ \mathcal{Z}_0 \\ \mathcal{M}\end{smallmatrix}\right]$ for that of $\left[\begin{smallmatrix}\mathcal{Z}_0 \\ \mathcal{M}\end{smallmatrix}\right]$. This aspect deserves further investigation.  We finally observe that the constraint 
\eqref{eq:SDP2-extended}
that appears in the case of nonlinear systems reduces to the constraint \eqref{eq:SDP2-extended-linear} in the case of linear systems. In other words, the program \eqref{eq:SDP-extended} is a 
nonlinear extension of the program \eqref{eq:SDP-extended-linear}.

\begin{example}\label{ex_linear}
Consider the thickness control system in a rolling mill \cite[Example 4.3]{isidori2017lectures}:
\begin{subequations}\label{sys}
\begin{align}
&\begin{bmatrix}
\dot x_1\\
\dot x_2
\end{bmatrix} = \begin{bmatrix}
 0 & 1\\
0 & 0
\end{bmatrix}\begin{bmatrix}
 x_1\\
x_2
\end{bmatrix} + \begin{bmatrix}
 0\\
b
\end{bmatrix} u \\
& h = \frac{1}{E+F} (E x_1 + FH + E d) \\
&  e  = h - h_r
\end{align}
\end{subequations}
where $x_1$ denotes the unloaded screw position, $u$ is the control input, $e$ is the error variable, $H$ and $h$ are the input and exit thickness, respectively, $h_r$ is the constant reference exit thickness and $d$ is the periodic disturbance induced by the eccentricity of rolls. Here, we assume that $d(t)=\sin(\sigma t)$. $E$, $F$, $H$, $b$ and $\sigma$ are constant parameters, where $E$, $F$, $H$, and $b$ are unknown and $\sigma$ is known. The control objective is to regulate the exit thickness $h$ to a desired constant reference $h_r$. Let $w_1 = h_r$, $w_2 = FH$,  $w_3 = \sin(\sigma t)$ and $w_4 = \cos(\sigma t)$, and then the error variable $e$ can be rewritten as
\[
e=C_e x+Q_e w
\]
where $C_e = \begin{bmatrix}
  \frac{E}{ E+F} & 0
\end{bmatrix}$ and $Q_e = \begin{bmatrix}
   -1& \frac{1}{E+F} & \frac{E}{E+F} &0
\end{bmatrix}$, and $w \in \mathbb{R}^4$ satisfies
\[
\dot w = {\scriptsize\begin{bmatrix} 0& 0& 0& 0\\0& 0& 0& 0  
\\0& 0& 0& \sigma\\0& 0& -\sigma& 0 \end{bmatrix}} w := Sw.
\]
Note that the minimal polynomial of $S$ is $m_S(\lambda) = \lambda^3 + \sigma^2 \lambda$, and thus the internal model \eqref{im-matrices} takes the form of
\[
\dot \eta = {\scriptsize\begin{bmatrix} 0& 1& 0\\0& 0& 1  
\\0& -\sigma^2& 0\end{bmatrix}} \eta + {\scriptsize\begin{bmatrix} 0\\0 
\\1 \end{bmatrix} }e.
\]
Setting the parameters as $E=1$, $F=2$, $H=1$, $h_r = 0.5$, $\sigma=1$ and $b=3$, we collect $T =10$ samples, i.e. $\mathbb D := \left\{ (x_i,u_i, e_i, \dot x_i, \eta_i )\right\}_{i=0}^{9}$,  by running an experiment with input uniformly distributed in  $[-0.1,0.1]$,
and with an initial state within the same interval. The reduced matrix $\mathcal{M}$ is given as 
\[
\mathcal{M} = {\scriptsize\begin{bmatrix}
1 &  1 & \ldots &  1\\ 
\cos(t_0)&  \cos(t_1) & \ldots & \cos(t_9)\\ 
\sin(t_0)&  \sin(t_1) & \ldots & \sin(t_9)
\end{bmatrix}}.
\]
Note that $\mathcal{M}$ has 3 rows after one redundant row associated with
the eigenvalue $\lambda = 0$ in the matrix $M$ is removed. The SDP is feasible and returns the controller gain
$\mathcal{K} = \begin{bmatrix}
-0.9966  & -0.5186  & -0.3485 &   0.2510 &  -0.7884
\end{bmatrix}$. The evolution of $x$, $h$ and $u$ with initial state uniformly distributed in  $[-1,1]$ are displayed in Figure \ref{linear}, and the closed-loop trajectories converge to a periodic solution of period $2 \pi$ where the regulated output is zero. \qedp

\begin{figure*}[htbp] \centering
\begin{minipage}{0.3\linewidth} \includegraphics[width=1\linewidth]{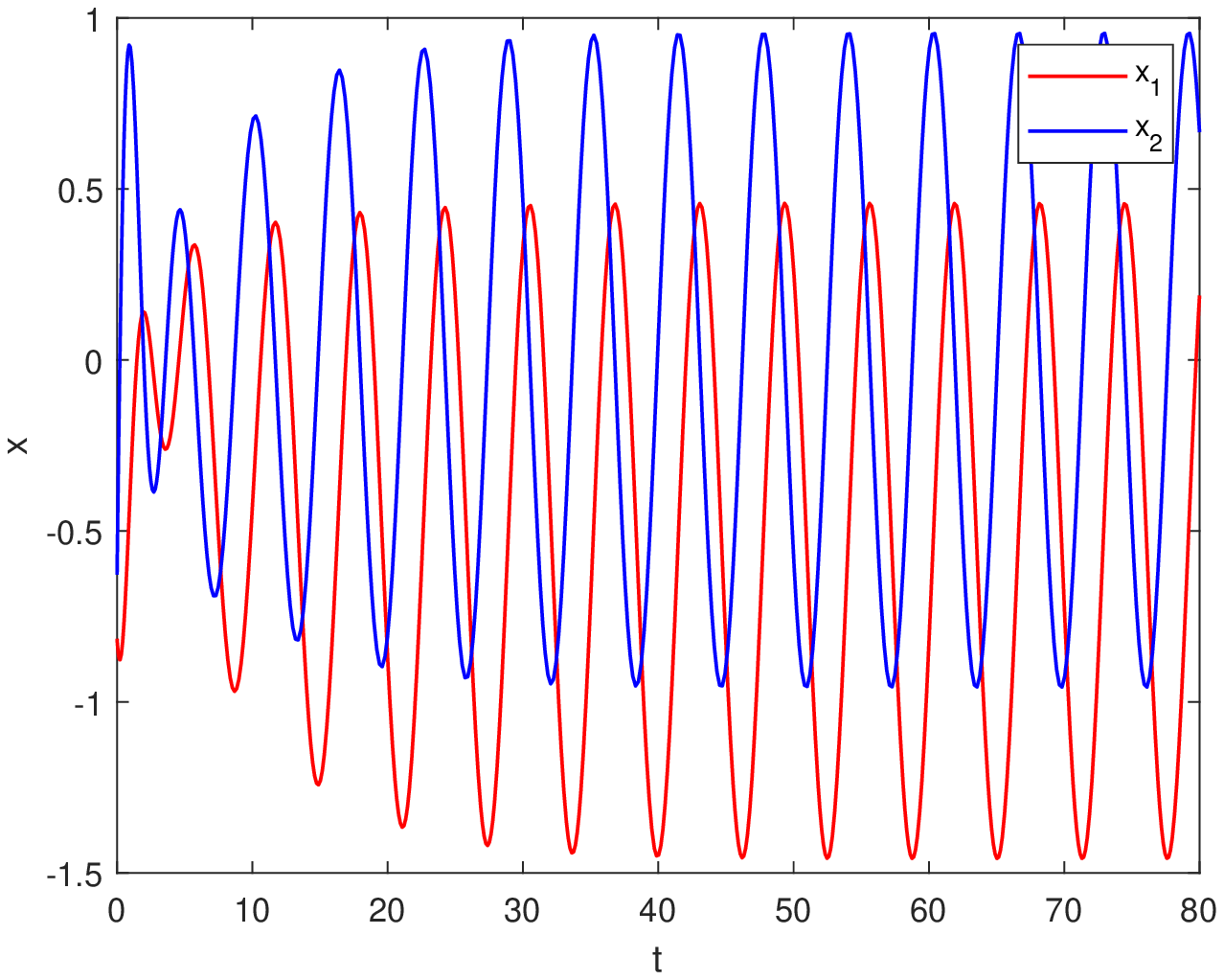}
\end{minipage} 
\begin{minipage}{0.3\linewidth} \includegraphics[width=1\linewidth]{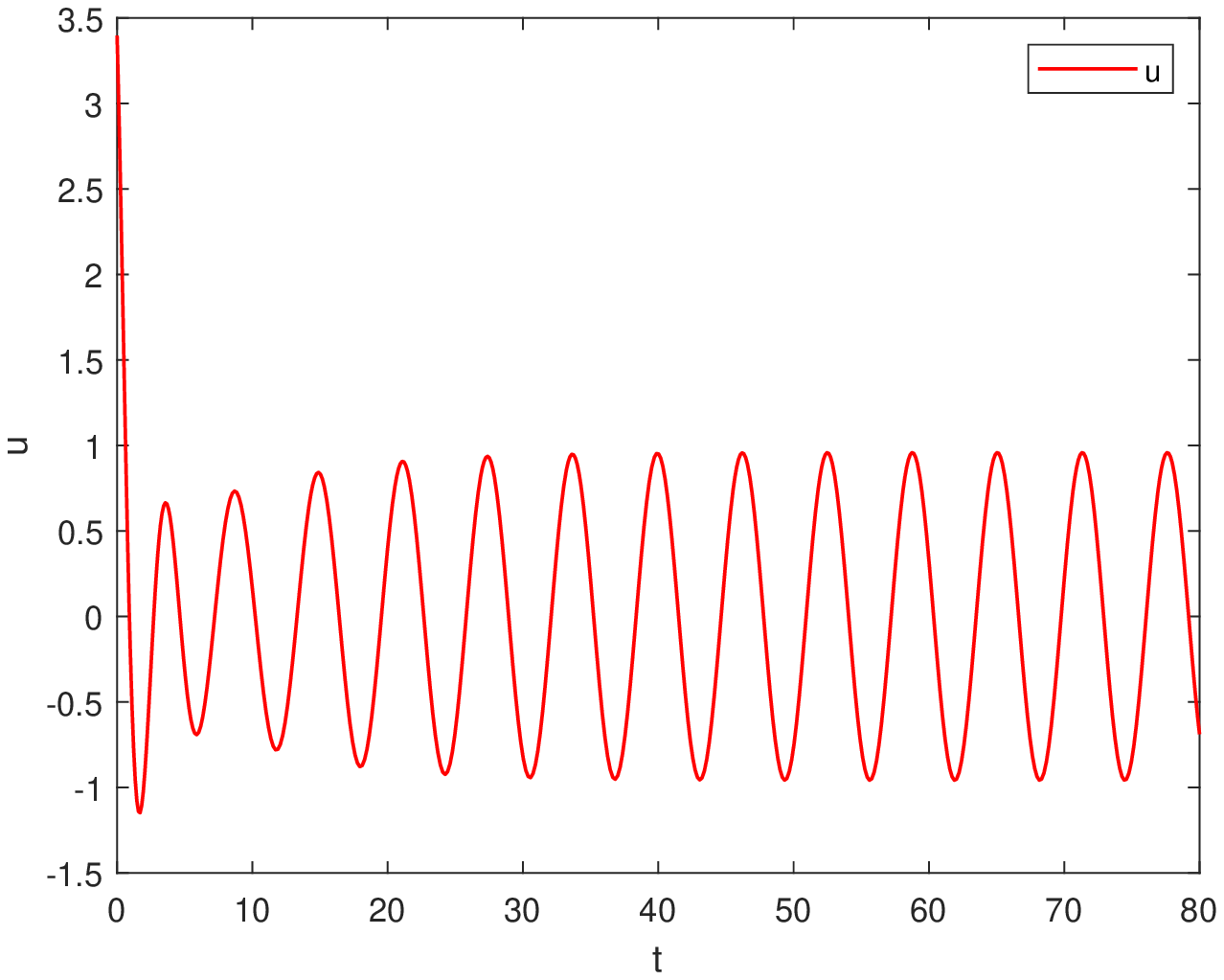}
\end{minipage} 
\begin{minipage}{0.3\linewidth} \includegraphics[width=1\linewidth]{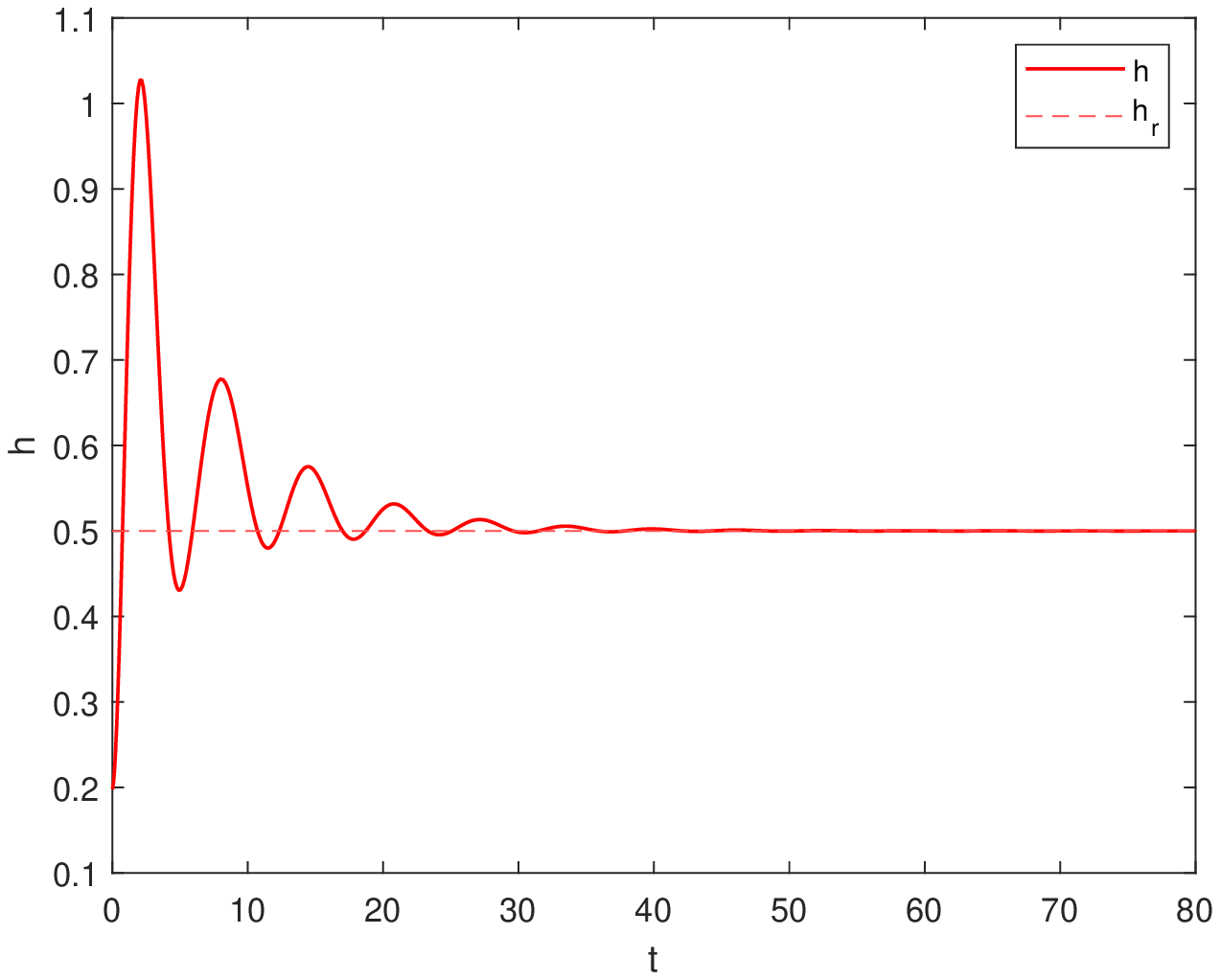}
\end{minipage} 
\caption{Evolution of $x$ (left), $u$ (center) and  $h$  (right) with initial state uniformly distributed in  $[-1,1]$.}\label{linear}
\end{figure*} 

\end{example}

\section{Conclusions and future work}

We proposed a data-driven design of controllers that solve 
{\color{black} a harmonic} 
output regulation problem for unknown nonlinear systems. {\color{black} Inspired by the model-based results of \cite{astolfi2015approximate,astolfi2022harmonic,giaccagli2024incremental},} we adopted an approach that does not require the existence of a normal form and privileges an approximate solution, which appears to be a natural viewpoint in the case the controlled plant is unknown. {\color{black}We have focused on continuous-time systems, but it is of interest to develop similar results for discrete-time systems. One reason to do so is to gain insights in the design of {\em output} feedback regulators, for which one could rely on results available for discrete-time systems.}  The breadth of the nonlinear output regulation theory suggests to explore other methods for the future, which might provide solutions with complementary features. Within the framework proposed in this paper, other aspects of interest are the use of projected dynamical systems to enforce constraints (see e.g.~\cite{lorenzetti2022pi}) and whether the noise filtering lemma can be extended to deal with unknown frequencies (see \cite{lorenzetti2022internal} for a recent original viewpoint on the tracking problem of exogenous signals of unknown frequencies {\color{black}and \cite{serrani2001semi} for a classical one}).

\bibliographystyle{elsarticle-harv} 

\bibliography{refs-4}

\appendix

\section{Proof of Lemma \ref{lem:forma-nuova-risposta-esosistema}}\label{app:proof}
Let $J= \mathcal{T}^{-1}S\mathcal{T}$ be the {\color{black} real} Jordan form of $S$, where
\begin{align*}
J = {\rm block.diag}\{J_{k_1} (\lambda_1), \ldots, J_{k_r} (\lambda_r), J_{k_{r+1}} (\lambda_{r+1}, 
\lambda_{r+1}^\star), \ldots\\, J_{k_{r+s}} (\lambda_{r+s}, 
\lambda_{r+s}^\star)\},
\end{align*}
$J_{k_i} (\lambda_i)$ is the Jordan block of order $k_i$ associated with the real eigenvalue 
$\lambda_i$, and  {\color{black}$J_{k_{r+i}} (\lambda_{r+i}, 
\lambda_{r+i}^\star)$ is the Jordan block of order  $k_{r+i}$ associated with the pair of complex eigenvalues $(\lambda_{r+i},  \lambda_{r+i}^\star)$} (hence, 
{\color{black}$J_{k_{r+i}} (\lambda_{r+i}, 
\lambda_{r+i}^\star)\in \R^{2 k_{r+i} \times 2k_{r+i}}$}). Note that  in $J$ above we are considering non-distinct eigenvalues, i.e., it can be that $\lambda_i= \lambda_j$ for $i\ne j$ and $\lambda_{r+i} = \lambda_{r+j}$ for $i\ne j$. 

The vector $\zeta = \mathcal{T}^{-1} w$ is correspondingly partitioned as 
$\zeta={\rm col}(\zeta^1, \ldots, \zeta^r, \zeta^{r+1},\ldots, \zeta^{r+s})$. Consider the evolution of the subvector $\zeta^i={\rm col}(\zeta^i_1, \ldots, \zeta^i_{k_i})$ associated with the Jordan block $J_{k_i} (\lambda_i)$. It is easy to show that the solution $\zeta^i(t)={\rm e}^{J_{k_i}(\lambda_i) t}\zeta^i(0)$ can be written as
\[
\zeta^i(t) = 
\underbrace{
\begin{bmatrix}
\zeta^{i}_{k_i}(0) & \zeta^{i}_{k_i-1}(0) & \ldots & \zeta^{i}_{2}(0) & \zeta^{i}_{1}(0)\\
0 & \zeta^{i}_{k_i}(0) & \ldots & \zeta^{i}_{3}(0) & \zeta^{i}_{2}(0)\\
0 & 0  &\ldots  & \zeta^{i}_{4}(0) & \zeta^{i}_{3}(0)\\
\vdots &\vdots &\ddots &\vdots&\vdots \\
0 & 0  &\ldots  & \zeta^{i}_{k_i}(0) & \zeta^{i}_{k_i-1}(0) \\
0 & 0  & \ldots  & 0& \zeta^{i}_{k_i}(0) \\
\end{bmatrix}
}_{=:L_{k_i}(w(0))}
\underbrace{
\begin{bmatrix}
\frac{t^{k_i-1}}{(k_i-1)!}{\rm e}^{\lambda_i t}\\[1mm]
\frac{t^{k_i-2}}{(k_i-2)!}{\rm e}^{\lambda_i t}\\[1mm]
\frac{t^{k_i-3}}{(k_i-3)!}{\rm e}^{\lambda_i t}\\
\vdots\\
t {\rm e}^{\lambda_i t}\\[1mm]
{\rm e}^{\lambda_i t}
\end{bmatrix}
}_{M_{k_i}(\lambda_i, t)}.
\]
Note that the notation $L_{k_i}(w(0))$ is used to indicate that the matrix corresponds to the block $J_{k_i}(\lambda_i)$ and that, instead of showing the dependence on $\zeta^{i}(0)$, the dependence on the initial condition in the original coordinates $w(0)$ is made visible. 

Similarly, consider the evolution of the subvector 
$\zeta^{r+i}={\rm col}(\zeta^{r+i}_1, \ldots, \zeta^{r+i}_{2k_{r+i}})$  associated with the Jordan block $J_{k_{r+i}}$ $(\lambda_{r+i}, \lambda_{r+i}^\star)$, where  $\lambda_{r+i},  \lambda_{r+i}^\star$ is  the pair of complex conjugate eigenvalues 
$\lambda_{r+i}= \mu_i+ \sqrt{-1} \psi_i,  \lambda_{r+i}^\star=  \mu_i- \sqrt{-1}\psi_i$. Then, \eqref{zeta} holds.  Hence, $\zeta(t)= L M(t)$, 
where  $L$ is the matrix in  \eqref{matrix-L}, 
\[\ba{l}
M(t):={\rm col} (M_{k_1}(\lambda_1, t), \ldots, M_{k_r}(\lambda_r, t),\\
\hspace{1.3cm} M_{k_{r+1}}(\lambda_{r+1},\lambda_{r+1}^*, t),
\ldots, M_{k_{r+s}}(\lambda_{r+s},\lambda_{r+s}^*, t))
\ea\]
and $w(t)=\mathcal{T}L M(t)$. This ends the proof. 
\qedp

\begin{figure*}
\be\label{zeta}
\zeta^{r+i}(t) = 
\underbrace{
\begin{bmatrix}
\begin{bmatrix}
\zeta^{r+i}_{2k_{r+i}-1}(0) & \zeta^{r+i}_{2k_{r+i}}(0)\\
 \zeta^{r+i}_{2k_{r+i}}(0) & -\zeta^{r+i}_{2k_{r+i}-1}(0)
\end{bmatrix}
&
\begin{bmatrix}
\zeta^{r+i}_{2k_{r+i}-3}(0) & \zeta^{r+i}_{2k_{r+i}-2}(0)\\
 \zeta^{r+i}_{2k_{r+i}-2}(0) & -\zeta^{r+i}_{2k_{r+i}-3}(0)
\end{bmatrix}
&
\ldots
&
\begin{bmatrix}
\zeta^{r+i}_{1}(0) & \zeta^{r+i}_{2}(0)\\
 \zeta^{r+i}_{2}(0) & -\zeta^{r+i}_{1}(0)
\end{bmatrix}
\\[3mm]
\mathbb{0}_{2\times 2}&
\begin{bmatrix}
\zeta^{r+i}_{2k_{r+i}-1}(0) & \zeta^{r+i}_{2k_{r+i}}(0)\\
 \zeta^{r+i}_{2k_{r+i}}(0) & -\zeta^{r+i}_{2k_{r+i}-1}(0)
\end{bmatrix}
&
\ldots
&
\begin{bmatrix}
\zeta^{r+i}_{3}(0) & \zeta^{r+i}_{4}(0)\\
 \zeta^{r+i}_{4}(0) & -\zeta^{r+i}_{3}(0)
\end{bmatrix}\\
\vdots &\vdots &\ddots &\vdots 
\\[3mm]
\mathbb{0}_{2\times 2}&\mathbb{0}_{2\times 2}&
\ldots
&
\begin{bmatrix}
\zeta^{r+i}_{2k_{r+i}-1}(0) & \zeta^{r+i}_{2k_{r+i}}(0)\\
 \zeta^{r+i}_{2k_{r+i}} (0)& -\zeta^{r+i}_{2k_{r+i}-1}(0)
\end{bmatrix}
\end{bmatrix}
}_{=:L_{k_{r+i}}(w(0))}
\underbrace{
\begin{bmatrix}
\displaystyle\frac{t^{k_{r+i}-1}}{(k_{r+i}-1)!}  {\rm e}^{\mu_i t}\cos(\psi_i t)\\[1mm]
\displaystyle\frac{t^{k_{r+i}-1}}{(k_{r+i}-1)!}  {\rm e}^{\mu_i t}\sin(\psi_i t)\\[1mm]
\hline
\displaystyle\frac{t^{k_{r+i}-2}}{(k_{r+i}-2)!} {\rm e}^{\mu_i t}\cos(\psi_i t)\\[1mm]
\displaystyle\frac{t^{k_{r+i}-2}}{(k_{r+i}-2)!} {\rm e}^{\mu_i t }\sin(\psi_i t)\\[1mm]
\hline
\vdots\\
\hline
{\rm e}^{\mu_i  t}\cos(\psi_i t)\\[1mm]
{\rm e}^{\mu_i  t}\sin(\psi_i t)
\end{bmatrix}
}_{=:M_{k_{r+i}}(\lambda_{r+i},  \lambda_{r+i}^\star, t)}
\ee
\end{figure*}

{\color{black}
\section{Periodic exponentially stable solution of the system \eqref{system-(9b)-(16)}}\label{AppB}
We focus on  the system \eqref{system-(9b)-(16)}, recalled below
\[
\begin{bmatrix}  
\dot x \\
\dot \eta
\end{bmatrix}
=
(\mathcal{A} +\mathcal{B} \mathcal{K})
\mathcal{Z}(x, \eta)+
\mathcal{P} w,
\]
and define it as
\be\label{f(z, w)}
\dot z = f(z, w),
\ee
where $z\in \R^{n_z}$ and $n_z=n+n_\eta$. 
Note that $f$ is continuously differentiable in $z$ (because so is $\mathcal{Z}(x, \eta)$) and smooth in $w$. We consider the class $\mathcal{W}$ of continuous inputs $w\colon \R\to \R^{n_w}$ that are bounded for all $t\in \R$. The inputs that satisfy Assumption \ref{assum:w} belong to $\mathcal{W}$. 

The goal of this appendix is to recall that there exists a solution $z_*(t):=(x_*(t),  \eta_*(t))$ of \eqref{f(z, w)} defined and bounded for all 
$t\in \R$, dependent on $w$,
which is globally 
 exponentially stable (hence, unique) and $D$-periodic {\color{black}(\ref{AppB})}. {\color{black}The appendix is added to make the article self-contained and contains no new results.} 
 
 We recall the following definition from \cite{pavlov2005convergent}:
\begin{definition}
The system \eqref{f(z, w)} is 
\begin{enumerate}
\item convergent if, for every $w\in \mathcal{W}$, there exists a solution $z_*^w(t)$ that is defined and bounded for all $t\in \R$, and is globally asymptotically stable. 
\item uniformly convergent if it is convergent and $z_*^w(t)$ is globally uniformly
asymptotically stable.
\item exponentially convergent if it is convergent and $z_*^w(t)$ is globally exponentially stable.
\end{enumerate}
\end{definition}

The following result holds \cite{pavlov2004convergent,pavlov2005convergent}:
\begin{theorem}\label{th:demid-pavlov}
 Assume that
there exist $P_1\succ 0, \beta>0$ such that for all 
$(z,w)\in \mathbb{R}^{n_z}\times  \mathbb{R}^{n_w}$
\[
\displaystyle\frac{\partial f(z, w)^\top}{\partial z} P_1^{-1}+P_1^{-1}
\displaystyle\frac{\partial f(z, w)}{\partial z}\preceq -\beta P_1^{-1}. 
\]
Then the system is exponentially convergent. In addition, if,  for a given $w\in \mathcal{W}$ and $D\in \mathbb{R}_{>0}$, $f(z,w(t))$ is $D$-periodic, then $z_*^w(t)$ is periodic of period $D$. If $f(z,w(t))$ is time-invariant, then $z_*^w(t)$ is constant. 
\end{theorem}

{\emph Proof.} Fix an input $w\in \mathcal{W}$ and a vector $\overline z$ such that $|f(\overline z, w(t))|\le \overline f<\infty$ for all $t\in \R$. Denoted by $V(z)$ the function $(z-\overline z)^\top P_1^{-1} (z-\overline z)$, it holds that $\dot V(z):=\frac{\partial V}{\partial x} f(z, w(t))<0$ for all $x\in 
\{z\in \mathbb{R}^n\colon V(z)> \gamma\}$ and all $t\in \R$, where $\gamma = (2 \beta^{-1} \bar f  \|P_1^{-1/2}\|)^2$. This implies that the set $\mathcal{R}_\gamma =\{z\in \mathbb{R}^{n_z}\colon V(z)\le  \gamma\}$ is a closed bounded set of $\mathbb{R}^{n_z}$ that is forward invariant  for  $\dot z=f(z,w(t))$. Hence, it  must  contain  at least one solution $z_*^w(t)$ defined for all $t\in \R$ \cite[p.~260]{pavlov2004convergent}. Actually, this solution is unique \cite[p.~261]{pavlov2004convergent}. 

Consider now a  solution $z(t)$ of $\dot z = f(z,w(t))$, $z(t_0)=z_0\in \R^{n_z}$, $t_0\in \R$, defined on its maximal interval of existence. Consider the function $V(z(t)-z_*^w(t))= (z(t)- z_*^w(t))^\top P_1^{-1} (z(t)- z_*^w(t))$ defined over the interval of  existence of $z(t)$. For every $t$ in this interval, it can be shown that $\dot V(z(t)-z_*^w(t))\le -\beta V(z(t)-z_*^w(t))$, hence $z(t)- z_*^w(t)$ is bounded and therefore $z(t)$ is bounded because so is $z_*^w(t)$. Then the interval of existence of $z(t)$ can be extended indefinitely, that is, $z(t)$ exists for all $t\in [t_0,+\infty)$. Then $V(z(t)-z_*^w(t))$ exists for all $t\in [t_0,+\infty)$ and satisfies
$\dot V(z(t)-z_*^w(t))\le -\beta V(z(t)-z_*^w(t))$, showing the exponential convergence of $|z(t)-z_*^w(t)|$ to zero.  Because $z(t)$ is a generic solution, we obtain that $z_*^w(t)$ is globally exponentially stable. We conclude that  the system \eqref{f(z, w)} is exponentially convergent. If the system \eqref{f(z, w)} is exponentially convergent, then it is uniformly convergent. By \cite[Property 3]{pavlov2005convergent}, the last part of the thesis follows. \qedp
}

\section{A few facts about the Fourier series}

{\color{black}In the proof of Theorem \ref{cor:integr-control} a few facts about the Fourier series of $e_*(t)$ are used, which we find it useful to summarize here. Although the function $e_*(t)$ is continuously differentiable, the facts will be given  under the assumptions that normally appear in textbooks such as \cite{tolstov2012fourier}. Moreover, we will use the generic symbol $v$, instead of $e_*$, for the function of which we consider the Fourier series.} 

The Fourier series of the $D$ periodic and {\color{black} absolutely} integrable on $[0,D]$ function\footnote{The function  $v$ is integrable on  $[0,D]$ if its integral on $[0,D]$ exists \cite[p.~8]{tolstov2012fourier}. If $v$ is continuous on $[0,D]$, then it is integrable on $[0,D]$. The function  $v$ is {\color{black}absolutely} integrable on  $[0,D]$ if the integral of its absolute value, 
{\color{black}$\int_0^{D} |v(t)| dt$,  
exists \cite[p.~8]{tolstov2012fourier}}.
} $v\colon \mathbb{R}\to\mathbb{R}$, is
\be\label{fourier.series}
\frac{a_0}{2}+\sum_{k=1}^\infty (a_k \cos (k \frac{2\pi}{D}t)+b_k \sin (k \frac{2\pi}{D}t))
\ee
where 
\be\label{coeff.fourier.series}
\ba{rll}
a_k =& \displaystyle\frac{2}{D} \int_0^{D} v(t)\cos (k \frac{2\pi}{D}t)dt, & k=0,1,2,\ldots\\
b_k =& \displaystyle\frac{2}{D} \int_0^{D} v(t)\sin (k \frac{2\pi}{D}t)dt, & k=1,2,\ldots
\ea
\ee
It is also worth recalling that the Fourier series \eqref{fourier.series} does not necessarily converge to $v(t)$, unless the following conditions hold {\color{black}\cite[Section 6, p.~75]{tolstov2012fourier}}:

\begin{theorem}\label{thm-pw-convergence}
{\rm ({\color{black}Convergence at continuity points})}
If the $D$-periodic function $v:\mathbb{R}\to\mathbb{R}$ is piecewise continuous in $[0,D]$ 
 and integrable
in $[0,D]$,
then 
at every point $t$ 
where $v$ is continuous and where both $\lim_{\Delta t\to 0^+} \frac{v(t +\Delta t)-v(t)}{\Delta t}$
and 
$\lim_{\Delta t\to 0^-}$ $\frac{v(t +\Delta t)-v(t)}{\Delta t}$ exist and are finite (in particular, where $v$ is differentiable, i.e., $\lim_{\Delta t\to 0^+} \frac{v(t +\Delta t)-v(t)}{\Delta t}=\lim_{\Delta t\to 0^-}$ $\frac{v(t +\Delta t)-v(t)}{\Delta t}$), 
the Fourier series \eqref{fourier.series} converges to $v(t)$.
\end{theorem}

If {\color{black}$v$ is a square integrable function on $[0,D]$, i.e., both $v$  and its square are integrable on $[0,D]$ \cite[p.~50]{tolstov2012fourier}},
then
Parseval's identity holds \cite[Chapter 5, Section 3, p.~119]{tolstov2012fourier}
\be\label{l2-norm-v}
\displaystyle \int_0^{D} v(t)^2 dt = \frac{D}{2} \left(\frac{a_0^2}{2} +\sum_{k=1}^\infty (a_k^2 +b_k^2)  \right),
\ee
{\color{black} and the Fourier partial sums 
\[
S_N(t)=\frac{a_0}{2}+\sum_{k=1}^N (a_k \cos (k \frac{2\pi}{D}t) +b_k \sin (k \frac{2\pi}{D}t))
\] 
converge to $v(t)$ in the $\mathcal{L}_2[0,D]$-norm sense, namely
\[
\lim_{N\to +\infty} \left(\frac{1}{D}\int_0^D (S_N(t)-v(t))^2 dt\right)^{1/2} =0
\] \cite[Theorem 2 in Chapter 5, Section 3, p.~119]{tolstov2012fourier}. }

{\color{black} It is also of interest to differentiate the Fourier series. To this end, we assume that $v$ is  a continuous function (hence, integrable on $[0,D]$) with an absolutely integrable derivative on $[0,D]$.
As a first consequence, we have \cite[Theorem in Chapter 3, Section 11, p.~84]{tolstov2012fourier}}  
pointwise convergence of the Fourier series associated with $v(t)$:
\be\label{v.as.a.fourier.series}
v(t)=\frac{a_0}{2}+\sum_{k=1}^\infty (a_k \cos (k \frac{2\pi}{D}t)+b_k \sin (k \frac{2\pi}{D}t)), \quad \forall t\in \R.
\ee
{\color{black}Then, by \cite[Theorem 1 in Chapter 5, Section 8]{tolstov2012fourier}, the Fourier series of $\dot v(t)$ can be obtained from the Fourier series of $v$ differentiating the 
latter
(the right-hand side of \eqref{v.as.a.fourier.series}) term by term. Hence,  
\be\label{series.fourier.derivative}
\sum_{k=1}^\infty (-a_k k \frac{2\pi}{D} \sin (k \frac{2\pi}{D}t)+b_k k \frac{2\pi}{D} \cos (k \frac{2\pi}{D}t))
\ee
is the Fourier series of $\dot v$. }
{\color{black} By Parseval's theorem \cite[Chapter 5, Section 3, p.~119]{tolstov2012fourier}, if $\dot v$ 
is square integrable on $[0,D]$} 
i.e., both $\dot v$  and its square are integrable on $[0,D]$ \cite[p.~50]{tolstov2012fourier},
{\color{black} then}
\be\label{l2-norm-dot-v}
\displaystyle \int_0^{D} \dot v(t)^2 dt = \frac{D}{2} \left(\displaystyle\frac{2\pi}{D}\right)^2 \sum_{k=1}^\infty (a_k^2 +b_k^2 )k^2.
\ee
{\color{black}If 
$\dot v$ is integrable on $[0,D]$
and $\dot v$ is bounded, then $\dot v$ is square integrable on $[0,D]$ \cite[Chapter 2, Section 4]{tolstov2012fourier} and \eqref{l2-norm-dot-v} holds. Similarly, if $v$ is continuously differentiable and $\dot v$ is bounded, then $\dot v$ is square integrable on $[0,D]$, because continuity of $\dot v$ implies its integrability on $[0,D]$, and \eqref{l2-norm-dot-v} holds.
}
The Fourier series of $\dot v$ allows the designer to provide an estimate on the $\mathcal{L}_2$ norm of $v$ in the case a bound on $\dot v(t)$ is known and the first $\ell+1$ elements of the Fourier series are zero, as we will remind in Lemma \ref{zero-fourier-coefficients} below. 

\begin{lem}\label{zero-fourier-coefficients}
\cite[Proposition 3]{astolfi2015approximate} The following statements hold:
\begin{enumerate}[label=(\roman*)]
\item Consider the system
\be\label{sub-im}
\dot \zeta =
\phi_k
\zeta 
+
N v, \quad \phi_k=\begin{bmatrix}
0 & \omega_k\\
-\omega_k & 0 
\end{bmatrix}, \quad \omega_k=k \frac{2\pi}{D}
\ee
where $\omega_k=k \frac{2\pi}{D}$, $k$ is a positive integer, $D$ is a positive real number, $N\in \mathbb{R}^2$ is a nonzero vector and $v\colon \mathbb{R}\to \mathbb{R}$ is a continuous periodic signal of period $D$.  Let $\zeta^*(t)$ be a periodic solution of period $D$ of \eqref{sub-im}, then the $k$-th coefficient of the Fourier series of 
$v(t)$ must be zero, that is 
\be\label{coeff-k-Fourier} 
\int_0^{D} {\rm e}^{i k \frac{2\pi}{D}t} v(t)dt =0,
\ee
where, for the sake of compactness, we are using $i$ for the imaginary unit instead of $\sqrt{-1}$. 
\item Consider the system
\be\label{sub-im-0}
\dot \zeta =
\gamma v, \quad \gamma\in \R\setminus\{0\}
\ee
where $v\colon \mathbb{R}\to \mathbb{R}$ is a {\color{black}continuous} periodic signal {\color{black}of period $D$}. 
{\color{black} Let} 
$\zeta^*(t)$  {\color{black} be a periodic solution} of period $D$ {\color{black}of \eqref{sub-im-0}}, then the $0$-th coefficient of the Fourier series of 
$v(t)$ must be zero, that is 
\be\label{coeff-0-Fourier} 
\int_0^{D} v(t)dt =0
\ee
\item Consider the Fourier series of $v(t)$ and suppose that its coefficients satisfy $a_0=0$ and $a_k=b_k=0$ for $k=1,\ldots, \ell$. Suppose that {\color{black}$\dot v\colon\R\to\R$} is a {\color{black}continuous}  periodic signal of period $D$ and $|\dot v(t)|\le v_d$ for all $t\in [0,D]$. 
Then
\[
\displaystyle \int_0^{D} v(t)^2 dt 
\le \frac{ D  v_d^2}{\left(\frac{2\pi}{D}\right)^2  (\ell+1)^2}.
\text{\hspace*{\fill}~{\tiny $\square$}}
\]
\end{enumerate}
\end{lem}

 
\noindent 
{\it Proof.} (i)~Define $z =\zeta_1^* +i \zeta_2^*$, where $i$ is the imaginary unit. $z$ satisfies the equation 
\[
\dot z = -i \omega_k z+\hat N v
\]
where $\hat N:= N_1 + i N_2$. Note that $N\ne 0$ implies $\hat N\ne 0$. Solving the differential equation and evaluating the solution $z$ at time $D$ returns
\begin{align*}
z(D)&= {\rm e}^{-i \omega_k D} z(0)+\int_0^{D} {\rm e}^{-i \omega_k (D-t)} \hat N v(t) dt \\ &= {\rm e}^{-i \omega_k D} z(0)+{\rm e}^{-i \omega_k D} \hat N \int_0^{D} {\rm e}^{i \omega_k t}v(t) dt. 
\end{align*}  
It is easily checked that, as $\omega_k$ is a multiple of $\frac{2\pi}{D}$ and $z$ is $D$-periodic, it follows that   ${\rm e}^{-i \omega_k D} z(0)=z(D)$ and ${\rm e}^{-i \omega_k D} \hat N =\hat N\ne0$, hence $\int_0^{D} {\rm e}^{i \omega_k t}v(t) dt=0$, which is \eqref{coeff-k-Fourier}. 
Bearing in mind the expression \eqref{coeff.fourier.series} of the coefficients of the Fourier series of $v$, the identity \eqref{coeff-k-Fourier} is equivalent to having $a_k=b_k=0$ as claimed. \\
(ii).~It is immediate.\\
(iii).~Since $\dot v(t)$ {\color{black}is continuous}
and $|\dot v(t)|\le v_d$ for all $t\in [0,D]$, $\dot v(t)$ is 
square integrable on $[0,D]$. Moreover,   
 if $|\dot v(t)|\le v_d$ for all $t\ge 0$, then  $\int_0^{D}\dot v(t)^2 dt \le D v_d^2$. Since $a_0=0$ and $a_k=b_k=0$ for $k=1,\ldots, \ell$, then, by  \eqref{l2-norm-v},
\[
\displaystyle \int_0^{D} v(t)^2 dt = \frac{D}{2}  \sum_{k=\ell+1}^\infty (a_k^2 +b_k^2)
\]
Similarly, by Parseval's identity for $\dot v$ (see \eqref{l2-norm-dot-v}), we have 
\[
\ba{rl}
D v_d^2 &\ge \displaystyle \int_0^{D} \dot v(t)^2 dt = \frac{D}{2} \left(\frac{2\pi}{D}\right)^2\sum_{k=\ell+1}^\infty (a_k^2 +b_k^2) k^2  
 \\ &\ge \frac{D}{2} \left(\frac{2\pi}{D}\right)^2  (\ell+1)^2\sum_{k=\ell+1}^\infty (a_k^2 +b_k^2)
\ea
\]
thus, 
\[
\displaystyle \int_0^{D} v(t)^2 dt = \frac{D}{2}  \sum_{k=\ell+1}^\infty (a_k^2 +b_k^2)  \le \frac{D v_d^2}{\left(\frac{2\pi}{D}\right)^2  (\ell+1)^2}
\]
which shows how the $\mathcal{L}_2$-norm of $v$ decreases as $\ell$ increases. 
\qedp

\end{document}